# The Greenland Telescope – Construction, Commissioning, and Operations in Pituffik

Ming-Tang Chen[1], Keiichi Asada[1], Satoki Matsushita[1], Philippe Raffin[1], Makoto Inoue[1], Paul T. P. Ho[1,2], Chih-Chiang Han[1], Derek Kubo[1], Timothy Norton[3], Nimesh A. Patel[3], George Nystrom[1], Chih-Wei L. Huang[1], Pierre Martin-Cocher[1], Jun Yi Koay[1], Cristina Romero-Cañizales[1], Ching-Tang Liu[1,4], Teddy Huang[1], Kuan-Yu Liu[1,2], Tashun Wei[1], Shu-Hao Chang[1], Ryan Chilson[1], Peter Oshiro[1], Homin Jiang[1], Chao-Te Li[1], Geoffrey Bower[1], Paul Shaw[1], Hiroaki Nishioka[1], Patrick M. Koch[1], Chung-Cheng Chen[1], Ranjani Srinivasan[1,3], Ramprasad Rao[1,3], William Snow[1], Hao Jinchi[5], Kuo-Chang Han[5], Song-Chu Chang[5], Li-Ming Lu[5], Hideo Ogawa[6], Kimihiro Kimura[6], Yutaka Hasegawa[6], Hung-Yi Pu[1,7], Shoko Koyama[1,8], Masanori Nakamura[1,9], Daniel Bintley[2], Craig Walther[2], Per Friberg[2], Jessica Dempsey[2,10], T. K. Sriharan[3,11], Sivasankaran Srikanth[11], Sheperd S. Doeleman[3], Roger Brissenden[3], Juan-Carlos Algaba Marcos[1,12], Britt Jeter[1], Cheng-Yu Kuo[13], and Jongho Park[14]

(1) Academia Sinica Institute of Astronomy & Astrophysics, Taipei 106, Taiwan
(2) East Asian Observatory, 660 N. A'ohoku Pl., Hilo, HI 96720, USA
(3) Center for Astrophysics | Harvard & Smithsonian, 60 Garden St., Cambridge, MA 02138, USA
(4) National Chung-Shan Institute of Science and Technology, N.300-5, Lane 277, Xi An Street, Xitun District, Taichung City 407, Taiwan
(5) National Chung-Shan Institute of Science and Technology, No.566, Ln. 134, Longyuan Rd., Longtan Dist., Taoyuan City 325, Taiwan
(6) Osaka Metropolitan University, Gakuen-Cho Sakai Osaka, Sakai 599-8531, Japan
(7) Department of Physics, National Taiwan Normal University, No. 88, Sec. 4, Tingzhou Rd., Taipei 116, Taiwan
(8) Niigata University, 8050 Ikarashi-nino-cho, Nishi-ku, Niigata 950-2181, Japan
(9) Department of General Science and Education, National Institute of Technology, Hachinohe College, 16-1 Uwanotai, Tamonoki, Hachinohe, Aomori 039-1192, Japan
(10) ASTRON, Oude Hoogeveensedijk 4, 7991 PD Dwingeloo, The Netherlands
(11) National Radio Astronomy Observatory, 520 Edgemont Road, Charlottesville, VA, 22903
(12) Department of Physics, Faculty of Science, University of Malaya, 50603 Kuala Lumpur, Malaysia
(13) Physics Department, National Sun Yat-Sen University, No. 70, Lien-Hai Road, Kaosiung City 80424, Taiwan
(14) Korea Astronomy and Space Science Institute, Daedeok-daero 776, Yuseong-gu, Daejeon 34055, Republic of Korea

e-mail: mtchen@asiaa.sinica.edu.tw

**Abstract**

In 2018, the Greenland Telescope (GLT) started scientific observation in Greenland. Since then, we have completed several significant improvements and added new capabilities to the telescope system. This paper presents a full review of the GLT system, a summary of our observation activities since 2018, the lessons learned from the operations in the Arctic regions, and the prospect of the telescope.

Keywords: Radio Telescope, Submillimeter Astronomy, Astronomical Instrumentation.

## 1. Introduction

The Greenland Telescope [1] is a 12-meter radio telescope designed for astronomical observations in submillimeter (sub-mm) wavelengths on a high Arctic site. Currently, the telescope is located at Pituffik Space Base (PSB)[1] in the northwestern corner of Greenland and has participated in the observing campaigns of the Event Horizon Telescope [2] (EHT) and the Global Millimeter-wave VLBI Array [3] (GMVA). The first scientific result featuring the GLT revealed a panoramic picture of the black hole and its jet at a 3 mm wavelength [4]. The GLT project was initially a collaboration between the Academia Sinica, Institute of Astronomy and Astrophysics (ASIAA) and the Smithsonian Astrophysical Observatory (SAO). In 2022, the collaboration was expanded to include the Danish partners, represented by the Department of Physics and Astronomy at Aarhus University (AU), Niels Bohr Institute at the University of Copenhagen (NBI), National Space Science Institute at the Technical University of Denmark (DTU), the University of Southern Denmark (SDU).

---
[1] Formerly known as Thule AB before April 2023.

The GLT was retrofitted from one of the prototype antennas for the Atacama Large Millimeter/Submillimeter Array (ALMA). The ALMA North America Prototype Antenna (the Prototype henceforth), after being thoroughly evaluated [5], was not deployed as one of the ALMA production units. In 2010, the National Science Foundation (NSF) issued a call for repurposing the Prototype. We responded to the call and proposed to deploy this telescope to a high-altitude site for conducting very long-baseline-interferometry (VLBI) in sub-mm wavelengths to image black-hole shadows. The NSF granted our proposition and awarded the antenna to SAO in 2011. By this time, we had been conducting site surveys of atmospheric opacity. From the surveying data, Summit Station in Greenland was identified as the future site and thus the location warranted the name of the Greenland Telescope[6, 7, 8] (GLT) project.

After preliminary testing at the Very Large Array (VLA) site, the Prototype was disassembled, and components were sent for retrofitting [9]. In early 2016, we completed a partial pre-assembly at a temporary location[2] near Norfolk, Virginia, to ensure that the new and critical components were

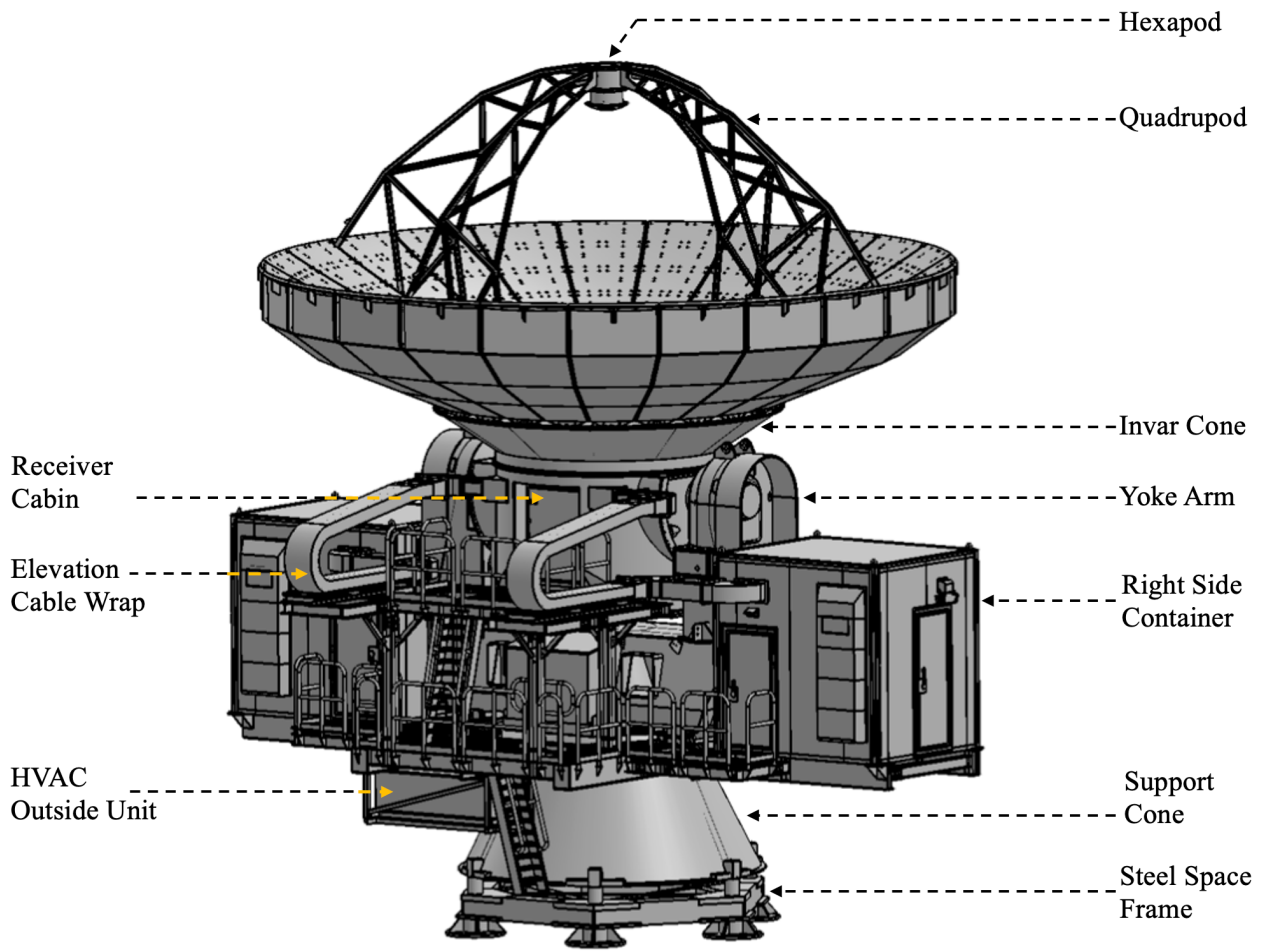

Figure 1. A graphical model of the GLT.

---

[2] World Distribution Services Chesapeake, 630 22nd St, Chesapeake, VA 23324.

Table 1
Antenna characteristics

| Keyword | Value |
| --- | --- |
| Primary Aperture | 12.0 m |
| Focal Length (Primary) | 4.80 m |
| fp / D of Primary | 0.40 |
| Secondary Aperture | 0.75 m |
| Final f / D | 8.00 |
| Primary Illumination (Edge Taper) | 12 dB |
| Magnification Factor | 20.0 |
| Primary Angle of Illumination | 128.02° |
| Secondary Angle of Illumination | 7.16° |
| Distance from Primary to 2nd Foci | 6.177 m |
| Depth of Primary | 1.875 m |
| Primary Vertex Hole Clear Aperture | 0.75 m |
| Elevation axis travel range | +1.89° to +90° |
| Azimuth axis travel range | -180° to +360° |
| Total weight | 112 Metric ton |

adequately fabricated. Afterward, most of the antenna components and the site trailers were shipped to Pituffik in the summer of 2016. The GLT was fully assembled in 2017 and detected its first radio signal by the end of the year [10, 11]. During the whole period, the National Chung-Shan Institute of Science and Technology (NCIST) has worked closely with ASIAA and contributed significantly to designing, testing, constructing components, and assembling the telescope. We also have engaged Vertex Antennentechnik (VA), the primary contractor for this antenna, since the testing in the VLA,

The location of PSB was chosen because it is the launching location for the Gree2nland Inland Traverse for transporting heavy components and equipment up to Summit Station. The PSB also provides a realistic Arctic environment during the winters for testing the newly assembled telescope, and significant engineering support exists on the base.

This paper presents an overview of the GLT system, including the antenna and the Pituffik site, the receiver system and the signal distribution networks, and the digital signal processing backends and data storage. It describes the techniques and procedures for optimizing the telescope's performance, such as improving the primary dish surface accuracy and the setup to closely monitor the signal stability, data fidelity, and quality. The GLT's on-sky performance and the scientific activities are covered in the later part of the paper, followed by the project's plan for the near future.

## 2. Overview of Antenna Construction

Deploying a 12-meter sub-mm telescope on Greenland's summit poses a tremendous technical challenge. The cold Arctic temperatures impose stringent requirements for operating the telescope. Instead of solid ground, the telescope will sit on a constantly shifting ice sheet, with the drifting snow threatening to bury any man-made structures. From the historical weather data of Summit Station[3], we have concluded three different weather conditions to define the GLT's operational specifications. Table 1 shows the antenna's mechanical/optical specifications to comply with the three weather conditions listed in Table 2. The primary operating conditions are weather conditions under which the full antenna performance should be achieved. These conditions cover 90 % of the recorded range in weather data. The secondary operating conditions cover the 90-95 % range for which a degraded antenna performance is accepted. The antenna will not be operated during the final 5 % of the time, where extreme weather conditions dictate a survival mode. The telescope will be in a stow position without sustaining damage to its structures and functions.

In Pittufik, the environmental temperature is generally warmer than at the Greenland summit. The GLT operations were taken care of by the new addition of the HVAC (heating, ventilation, and air conditioning). But historically[4], Pituffik has a much higher wind-speed record which might cause potential issues for the telescope and its observations. We have been closely monitoring the telescope's dynamic movement using a set of tiltmeters (see Section 8.5) mounted on the telescopes. So far, we have not experienced any detrimental effects due to the wind.

### 2.1 Preparation and New Elements

After the testing and disassembly of the Prototype at the VLA site in New Mexico, the antenna components were categorized into 3 types: to be re-used for the GLT without modification, to be reused after modifications, and to be discarded and re-designed with new parts made. The reflector backup structure (BUS), consisting of 24 segments, has been thoroughly inspected and repaired. The process has eliminated a large number of voids in the BUS structure. The quality of the new segments has benefited from the latest ALMA improvements for the Production Antennas. The elevation shafts and elevation bearings are the main components that were modified after disassembly at the VLA. The new components include the support cone, azimuth bearing, cabin HVAC system, whole servo system, all cables, quadrupod, and hexapod. In addition, we designed a primary mirror panel de-ice system and side containers to house the servo cabinets. Figure 1 presents a graphical illustration of the GLT.

#### 2.1.1 HVAC

It is essential to keep the structure of the receiver cabin at a stable temperature to avoid potential BUS deformations and associated problems. At the apex hole, the cabin is protected from weather by a mechanical shutter and sealed off from airflow with a Goretex (RA7956) membrane [12] of 0.5 mm in thickness. The shutter consists of 2 flaps, each covering half

---

[3] https://gml.noaa.gov/dv/site/sum/met.html.

[4] https://weatherspark.com/.

Table 2
Required operating conditions on Greenland's Summit.

| Keyword | Primary | Secondary | Survival |
|---|---|---|---|
| Ambient temperature | 0 to -50 °C | -50 to -55 °C | -55 to -73 °C |
| Vertical temperature gradient* | +12K to -1K | +15K to -2K | N/A |
| Wind speed | 0 to 11 m/sec | 11 to 13 m/sec | 55 m/sec |

*: The historical data of Summit Station indicate that during the winter time we would expect a temperature inversion layer, a vertical temperature gradient over the entire antenna height. A linear temperature distribution over the entire antenna height and, in particular, over the antenna's primary mirror shall be assumed. The numbers above are variations over the entire antenna height and are based on temperature measurements at 1m, 10m, and 20m above ground. A temperature gradient of + 12 K indicates a total temperature increase of 12 K from bottom to top of the antenna, and -1 K indicates a total temperature drop of 1K from bottom to top. The temperature inversion occurs during artic winter when there is no solar radiation and the atmospheric conditions are most stable and quiescent. From the same set of data, there is no noticeable temperature gradient in the summer months.

of the apex aperture and supported by a frame sitting on the invar cone. Each flap is activated by a motor to open or close the shutter. In addition, when the shutter is closed, an airflow blowing from the cabin to the outside, prevents snow accumulation on the shutter.

The new HVAC maintains the *receiver cabin* at a stable temperature by regulating the temperature of the cabin's air and walls, including the ceiling, a part of the invar cone for the BUS. Hence, it has adopted a means combining air circulation and a wall-tempering system.

VA provided a thermal study, which detailed the HVAC specifications and insulation requirements adapted to winter and summer worst-case scenarios at Summit Station. The study gave the following results for two worst-case scenarios in winter and one in summer. They are:

Case-1: When the outside air temperature is -55°C with a 13 m/s wind speed, and no Sun, while inside the receiver cabin, the temperature stays above 15°C. The resultant heat loss is 3.5 kW.
Case-2: In the survival mode with an outside air temperature of -73°C and a 55 m/s wind for the same receiver cabin temperature. The resultant heat loss is 4.2 kW.
Case-3: When the outside is 0°C with no wind, combined with solar irradiation of 800 W/m$^2$ on one side of the cabin and snow on the other. The result is that the antenna requires 4 kW of cooling.

The heat losses in winter can be compensated with internal heating in the receiver cabin. An outside unit under the antenna platform will provide the cooling power for summer.

In addition, to maintain the temperature homogeneity in the receiver cabin from +15 to +22°C ±1°C and to ensure the antenna's pointing accuracy, a tempering system is rebuilt for the cabin and the elevation shafts. A network of copper pipes with flowing coolant runs vertically along the walls of the receiver cabin, along the elevation shafts, and around the cylindrical part of the invar cone. Menerga [5], a VA's subcontractor, installed and tested the complete tempering system. Two indoor racks recirculate and filter the air and the glycol distribution unit. The system was partially tested in Pituffik for temperature reaching -40°C with the wind speed approaching the survival conditions.

Based on the thermal study results, we covered all parts of the antenna, except the dish surface and the quadrupod, with high-quality thermal insulation: 3-inch-thick elastomer foam for cryogenic, guaranteeing flexibility way below the GLT's survival temperature and providing a thermal conductivity of 0.034 W/(m·K) at -50°C and 0.028 W/(m·K) at -100°C. In addition, a thin layer of aluminum sheet enclosed the antenna to protect the foam and prevent moisture ingress.

### 2.1.2 Hexapod

The original hexapod mount built by VA for the Prototype did not meet the specifications for operations in cold temperatures (-55°C) at Greenland and could not be refurbished. For a replacement, ADS International has delivered a new hexapod compliant with GLT requirements. The new hexapod comes with the X & Y translations orthogonal to the optical axis and the Z translation parallel to the optical axis. The Hexapod system is equipped with a chopping function, the range is 600", with a 1 Hz frequency (measured) value with an accuracy better than 0.9" and a settling time of 0.3 s. The hexapod has survived tests at -70°C. It has been tested at -55°C in a chamber, and no degradation in range and speed has been observed. The hexapod is envisioned for single-dish operations. Since typical VLBI observations do not require a nutating secondary, the chopping function of the hexapod has not been thoroughly tested. Table 3 lists the major specifications of the new hexapod.

---

[5] https://www.menerga.com/ .

### 2.1.3 De-Ice

Ice or snow accumulation on the primary dish degrades the telescope's performance. It will increase the system noise temperature, and affect pointing, aperture efficiency, and beam shape (see Section 8.4), and might also change the polarization isolation. This is an essential issue for telescope operations in the polar environment. Following the experience from the South Pole Telescope [13], we have implemented a de-ice system to prevent rime ice and snow accumulation on the primary dish. The de-icing system is designed to raise the temperature of the dish panels by roughly 2°C above ambient temperatures. It does not intend to melt the accumulated snow but provides a physical process to speed up snow sublimation.

In planning the de-ice system, we divided the primary dish into four quadrants, each of which is further equally divided into three zones (similar to an equally-divided, 12-slice, round pizza). We use the primary 480 VAC 3-phase system for the heating and apply one of the three phases (277 VAC Line-to-Neutral) to each zone. The heating system is programmable. We may heat individual quadrants or in a sequence of order. Typically, the dish is heated cyclically through each quadrant until the panels reach the set temperature, nominally 2°C above the air temperature at the dish periphery. This rotation is for the initial phase only. Once the panels are at temperature, the heating will be sporadic, as needed to maintain the set temperature, with a hysteresis of about 0.4 K.

We have operated the de-ice system in the past three winters in Pituffik. Empirically, it helps clear the dish of snow and rime ice during the dark time of the year. We usually turn on the de-ice a few hours before an observation campaign but turn it off during the observation. When there is sunlight, the solar radiation has enough power to heat the dish and clear the snow. The de-ice functions will be ultimately tested at Summit Station. More on the de-ice will be discussed in Section 8.4.

### 2.1.4 Side Containers

In the original design, the Prototype housed its power distribution and other electrical components in the antenna support cone. It accommodated the servo system on its platform at the back of the antenna, exposed to the weather. Such an arrangement would not work for a telescope operating in the Arctic. Hence, we designed a set of container housings to accommodate most electronics: the servo cabinet and antenna control unit, hexapod control unit, side container HVAC (one unit in each container), receiver cabin HVAC control rack, de-ice controller, and power distribution cabinets. The side containers are equipped with their HVAC systems which are independent of that of the receiver cabin. The side containers and their support structure were built in Taiwan, and a blank assembly was performed at NCIST. The side containers were shipped to Pituffik in 2017 and integrated with the antenna for the first time.

Table 3
GLT hexapod sub-reflector characteristics

| Keyword | Value |
|---|---|
| Min. dynamic loading | 53 kg |
| Dimensions | Dia. 580 mm, Ht. 613 mm |
| Total weight | 67 kg |
| Interface material | Invar |
| Movement in x and y axis | |
| -   Distance | +/- 5 mm from center |
| -   Slewing Mode | Less than 0.5 mm/sec |
| -   Collimation mode | Min. steps of 20 μm, 10 μm differential accuracy |
| Movement in z axis (optical axis) | |
| -   Distance | +/- 10 mm from center |
| -   Slewing Mode | Less than 0.5 mm/sec |
| -   Focus adjustment | Min. steps of 20 μm, 5 μm differential accuracy |
| -   Focus switching | Perform cycles up to +1.5 mm and -1.5 mm with repeatability of +/- 20 μm |
| Nutating | |
| -   Chopping Range | 10 arcmin maximum |
| -   Accuracy | 1.3 arcsec |
| -   Direction | Any directions |
| -   Chopping Freq | 2 Hz |
| -   Chopping mode | Two-position chopping |
| -   Settling time | Less than 50 ms |

We decided to attach the containers to both sides of the antenna to minimize structural imbalance. The pointing and operations of the GLT are susceptible to the structural balance because the antenna will reside on the moving ice sheet of the Greenland summit. If the containers were at the back locations, the added weight would create too much imbalance affecting the antenna performance on an ice foundation. In addition, the antenna tilting would be a constant concern.

Nevertheless, The overturning moment due to the platforms and side containers is estimated to be 227 kN·m front to back. The value is well within the required static raceway safety of the azimuth bearing, and well within the original ALMA specifications [6]. We did not compute the maximum allowable unbalance torque, but even with 180 km/h winds and gusts up to 220 km/h, our safety factor is higher than 3.

### 2.2 Pre-Assembly in Norfolk, Virginia

---

[6] Based on the proprietary report provided by VA, "ALMA specification and Statement of Work Azimuth Bearing." we achieved a safety value above 5.

We assembled the antenna up to the elevation axis, including the receiver cabin, and tested all antenna parts, except the primary dish. Assembly started in early 2016 at World Distribution Services premises in Norfolk, Virginia. The activities were concluded with the shipment of four shrink-wrapped, out-of-gauge assemblies and all other components in standard 30-ft containers, including the dish, the quadrupod, and all the cables, at the end of June 2016.

One primary task of the pre-assembly was to assemble and align the receiver cabin, yoke arms, and elevation bearing. After assembly, we found that the cabin was initially off by 3.5 mm along the elevation axis. It was fixed by shifting the cabin with shimming along the elevation shaft. The azimuth axis verticality was 14.2", within the tolerable 20". The center point of the azimuth bearing to the vertical axis was 0.1 mm; the orthogonality between axes was 2", which was within the acceptable limit. The center point of the elevation axis to the vertical axis was within 0.4 mm, less than the maximum misalignment permitted of 0.5 mm. This newly formed unit was shipped as one unit.

## 2.3 Telescope Installation at Pituffik Space Base

After the antenna shipment arrived, we started erecting the antenna on a preset foundation at the GLT site. From July to September 2016, we repeated what we did in Norfolk and assembled the antenna and the receiver cabin up to the elevation axis level. In parallel with the Virginia activities, the primary dish was assembled in an unheated PSB facility. To achieve the best surface accuracy, it is required to assemble the BUS and the dish panel at a temperature close to that of telescope operations. Hence, we can only assemble the dish in Pittufik's winter time. By early 2017, the primary dish was assembled on the invar cone, with all the de-ice heaters installed. It was then mounted to the antenna in the early summer, followed by installing the power, electrical and electronic networks. With the antenna powered up, VA's engineer began the antenna servo commissioning in September, which lasted six weeks. The installation and testing of the receivers and associated electronics commenced in Nov., and by the end of 2017 Christmas, we detected the first radio signal from the Moon at 86 GHz. Figure 3 shows a nearly completed GLT in the winter of 2017.

### 2.3.1 Backup Structure Assembly

Before arriving at Pittufik, the BUS was assembled at the VA facility in Germany. It was a painstaking process to adjust and align the BUS to a satisfactory state. Afterward, every bolt and calibrated spacer connecting the segments was labeled and recorded. In Pittufik, the segments were precisely mounted on the invar cone following the record of the pre-assembly result. In addition, we also mounted the quadrupod and secondary housing (the head part) onto the dish. The process took four weeks. Figure 2 shows the assembled BUS in the unheated indoor facility in PSB.

In the spring of 2017, the assembly work continued with mounting the de-ice heater on each dish panel, followed by the panel installation onto the BUS. The later step required accurate alignment of every panel. The job was contracted to SETIS, a French geodetic company with notable experience in ALMA panel alignment. The surveyor used a laser tracker and reached a 40 μm root-mean-square (RMS) error over the assembled dish surface. This value was calculated from the deviations of the 1320 points, i.e., 5 points per panel for all 264 panels, from a perfect paraboloid dish.

After assembly at the site, the first photogrammetric survey (see Section 6) showed the dish surface warped slightly into a saddle-type pattern. Its surface accuracy worsened to 180 μm RMS. The degradation in the dish surface accuracy can be attributed to two primary reasons:

- The sixteen outermost panels need to be removed to allow the mounting of the lifting lugs for the dish.
- The BUS must be removed from its support: the invar cone for any lifts.

It was not until the summer of 2018 that we could improve the dish surface accuracy after the photogrammetry campaign. Further details about the dish measurements are given in Section 6.

### 2.3.2 Antenna Foundation

The GLT site in PSB was chosen to have a clear view of our interested science targets, with easy access to power and communication resources on the base. We used the active runway heading as a reference for the True North heading. The antenna mounting system employs six (6) equally spaced hard points. We used a theodolite to place these hard points equally spaced about True North. At the GLT site, the telescope foundation was completed before the shipment arrived from Norfolk in mid-July 2016. The foundation consisted of a mean 30-cm leveled layer of asphalt on the tarmac applied in two iterations on which wood pads were set, providing the matching interface to the Antenna support base. As shown in Figure 1, there is a steel space frame underneath the antenna support cone, which is sitting on the wood pads (not shown) to better distribute the antenna load. The steel frame reinforces the antenna stability during operations and to guarantee pointing accuracy, while the wood pads served as a thermal barrier between the antenna and the ground. The wooden pads were made of cross-laminated Douglas Fir timbers, each 75 inches square and 8 inches thick. The interfaces between the wood pads and the steel foundation mounting feet were drilled, bolted, torqued, and grouted for better load distribution. The foundation was leveled within 2" to 5" in all planes, but its alignment to True North was off by 6°, found later from the pointing observations.

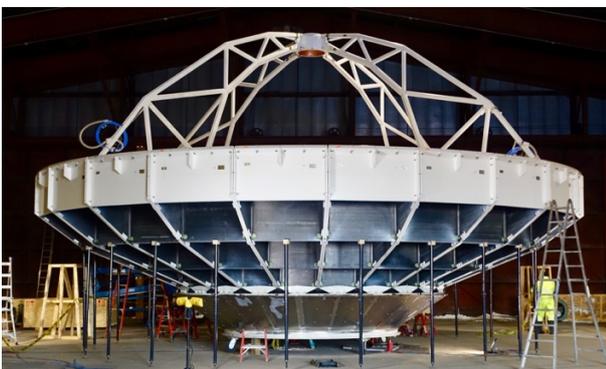

Figure 2. The backup structure is shown after the assembly of the carbon fiber segments and quadrupod on the invar cone in one of the hangars of the PAB.

After the ship arrived, we successfully mounted the support cone with the azimuth bearing, the yoke traverse, and the receiver cabin/yoke arms unit. The support cone was aligned with an accuracy of 0.01° in the N-S direction and better than 0.01° in the E-W direction. The following 2 months were spent outfitting the telescope mount with mechanical equipment, electronics, and thermal insulation. By the fall of 2016, the antenna was turning in azimuth and elevation.

*2.4 Maintenance*

VA has documented the GLT Mechanical Subsystem's maintenance in a Technical Note, followed by all tasks carried out on the subsystems needing care. This maintenance is executed periodically, as recommended, by GLT technicians and engineers or/and TAB personnel.

A complete lubrication with oil changes is done every three years on the azimuth and elevation gearboxes. This primary maintenance also includes greasing of azimuth and elevation bearings. Toothing on the elevation rims has to be greased more frequently and at least once a year, as the teeth are exposed to the environment. Azimuth-bearing teeth are enclosed, and grease does not dry as fast. Other miscellaneous tasks are performed regularly, such as inspecting, checking, and fixing bolts, including bolts for the outer platforms.

General maintenance does not require significant parts dismounting, but the elevation gearboxes' maintenance is the most cumbersome. We can only reach their protection panels by opening up heavy accessing doors from inside each side container. A bolted cover also protects the azimuth bearing, and access to the teeth requires removing steel protection panels. Azimuth-bearing lubrication is easily performed from inside the support cone.

## 3. Instrumentation

The first-light instrument for the GLT is a three-cartridge receiver system. Its cryostat has a similar design to the dewar

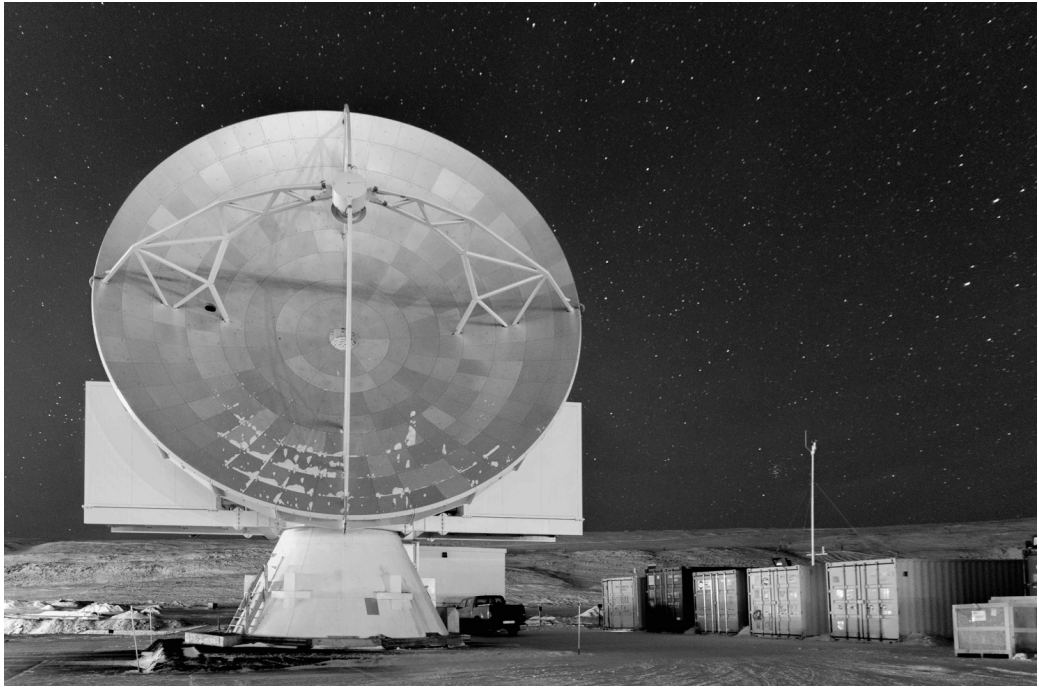

Figure 3. The GLT under commissioning in PSB [10]. The newly added side-containers are partially seen behind the primary dish. The containers shown in the lower right corner of the photo are our working spaces and storage facilities. Some snow can be seen in the lower part of the primary dish. At that moment, the de-ice system was not yet implemented. The hole on the left side of the dish, near the base of the quadrupod, is the location of the optical guide scope.

Table 4
The GLT first-light receivers

| Keyword | Rx86 | Rx230 | Rx345 |
| --- | --- | --- | --- |
| Origin | ASIAA | OPU | IRAM |
| Polarization | Dual Circular | Dual Circular | Dual Linear |
| RF freq. range | 84 – 96 | 213 – 231 | 275 – 373 |
| LO freq. range | 80 – 88 | 221 – 223 | 283 – 365 |
| 1st stage detector | MMIC | SIS | SIS |
| Mixing scheme | USB after 1st mixer | 2SB | 2SB |
| # of IF channel | 2 | 4 | 4 |
| IF (GHz) | 4 – 8 | 4 – 8 | 4 – 8 |
| VLBI LO freq. (GHz) | 80.6 | 221.1 | 342.6 |
| $T_{SSB}$ (K) 80% LO range (All) | 90 | 70 (110) | 147 (219) |

The abbreviations: Radio Frequency (RF); Intermediate Frequency (IF); Local Oscillator (LO); Two Sideband (2SB); Upper Sideband (USB); Monolithic Microwave Integrated Circuit (MMIC); Superconductor-Insulator-Superconductor (SIS).

used for a Japanese project [14]. It accommodates three ALMA-type receivers or any cartridges with the same mechanical interfaces as the ALMA's. The GLT cartridges currently are an 86 GHz cartridge of our design (Rx86), a 230 GHz cartridge from Osaka Prefecture University (OPU) [15] (Rx230), and an ALMA Band 7, 345 GHz cartridge from Institut de Radioastronomie Millimétrique (IRAM) [16] (Rx345). While currently exclusively used for VLBI, these receivers can also be used for single-dish observations. We purposely developed the GLT system with the same or similar interfaces as the ALMA to host ALMA-type receiver cartridges with minimum modifications. A detailed account of this receiver can be found in Han et al [17].

Figure 4 shows a block diagram of the GLT receiver system. The diagram illustrates the configuration of the three receiver cartridges and the connections with their associated warm parts. Table 4 lists the critical specifications of the GLT receiver system. The IF/LO signal distribution will be described in Section 4.

### 3.1 The Cryostat

The cryostat uses a three-stage, Gifford-McMahon (GM) cryocooler[7], which is specified to deliver 33 W of cooling power at 80 K (nominally), 8 W at 15 K, and 1 W at 4 K. Since this cold head is designed to support the ALMA cryostat with ten cartridges, its cooling power is rather excessive for our three-cartridge system. The ALMA and GLT cryostat both has a 110-K stage where the infrared (IR) filters are mounted. Based on the ALMA specifications[8], we assume that the heat load due to the dissipation of electronic components mounted on the 4 K stage to be 72 mW, equally distributed among two high-frequency cartridges. The total loading due to conduction by wiring should be 10 mW. For all three cartridges, the total heat loads on the 15 K and 110 K stages should be less than 500 mW and 4 W, respectively.

The cryostat is mounted directly to the primary dish structure, and thus it changes its orientation with the antenna pointing. As the antenna tips toward the horizon, the cold head generally loses a few percent of cooling power, which does not cause major issues for the operations. This model usually delivers more cooling power in an upright operational position. The cold head comes with a 2-liter cryogen buffer tank for maintaining its cooling stability and uses an air-cool compressor to power its functions.

Similar to the ALMA front-end configuration[18, 19], the GLT cryostat is concentrically aligned with the antenna's optical axis. It is mounted with the Cassegrain focus lying at the center of the cryostat's top plate. The top plate has three distinctive circular metal plates; each mounted with a vacuum window specific to one of the receiver bands. For installation, the receiver cartridges are inserted into the bottom of the cryostat via guided rails.

### 3.2 Rx86 Cartridge

Each receiver cartridge has its associated optics to refocus the incoming signal from the antenna. The Rx86 optics comprises a pair of plano-ellipsoidal reflectors on the cryostat. The optics folds and guides the incoming signal into the cryostat through a Teflon vacuum window. Once inside, the signal passes through two IR filters located at the 110 K and 15 K thermal shields, respectively, and arrives at a phase correction lens at the aperture of the feed horn. The feed horn and lens combination is the same design as ALMA Band-3 [20].

Following the feed horn is the RF front-end section, which comprises a W-band phase shifter[9], a short section of the interfacing waveguide, and an orthomode transducer (OMT).

---

[7] Sumitomo Heavy Industries, Ltd. Cryogenics Division, Tokyo, Japan.

[8] ALMA Project, Cryostat Technical Specifications, FEND-40.03.00.00.B.SPE, 2003-09-20.

[9] Sivasankaran Srikanth, NRAO, private communication.

The phase shifter is an electroformed, square WR-10 waveguide of one inch in length. It has internal grooving on one pair of opposite surfaces. The waveguide is made of copper with Au coated, providing $90°\pm2°$ phase offset from 85–95 GHz, and $94°\pm4°$ degrees from 77–97 GHz. Insertion loss is 0.8 dB.

The OMT has an insertion loss of less than 0.2 dB, a return loss of around 20 dB, and more than 40 dB isolation over the 75–110 GHz band. After the OMT, each polarized channel passes through two cascaded amplifier-isolator chains. Each amplifier contains a four-stage, InP high electron mobility transistor (HEMT) fabricated as a monolithic millimeter-wave integrated circuit. Cooled to 20 K, its noise temperature is typically 60–80 K, with power gain higher than 22 dB over 80–105 GHz.

The RF section is thermally anchored to the 15 K stage of the cryostat. After the isolator, the signal chain is connected to a high-pass filter followed by a double-balance mixer located at the 110 K stage. The high-pass filter, designed with a cutoff frequency of 84.5 GHz, is combined with the mixer to achieve an upper-sideband conversion for this receiver channel. The mixer is commercially available[10] and requires a nominal driving power of 12 dBm for maximum IF output performance. The cartridge delivers two upper-sideband (USB) IF signals of orthogonally circular polarization. The design of the entire receiver cold section is adapted from a previous project [21] [22]. For the convenience of discussion, we will, henceforce, refer to the two polarization states as POL-0 and POL-1 regardless whether they are linear or circular polarized.

Once exiting the cold cartridge assembly (CCA), the two IF signals are connected to the warm cartridge assembly (WCA). For the Rx86, its WCA is a modified version of the ALMA Band-6 WCA [23]. The IF signals are amplified, conditioned, and then sent out of the receiver system for further processing. The WCAs are the modules that synthesize the LOs for the CCA front end. More details about the WCA modules will be discussed in the later sections.

### 3.3 Rx230 Cartridge

The Rx230 was procured through collaborative efforts with the National Astronomical Observatory of Japan and OPU. The receiver is designed to detect dual circular polarized signals at 221 GHz [24]. Its front end is based on superconductor-insulator-superconductor mixer (SIS) technology and, thus, must be thermally anchored to the 4 K stage of the cryostat. The incoming signal enters the cryostat through an anti-reflection quartz window and passes through two IR filters at 110 K and 15 K, respectively. The signal is refocused through a pair of cold optics before entering the feed horn, and the entire optics plus the front end are tightly packed on its 4 K platform. The circular polarizer is a septum-type waveguide that operates from 210 GHz to 245 GHz [25]. It has a transmission loss of about 1.5 dB at room temperature and a cross-talk level better than -20 dB. We use an ALMA Band-6 WCA as the IF/LO module. This cartridge puts out four IF signals from two polarization channels, with each producing two IF signals of the upper and lower sideband (USB and LSB).

### 3.4 Rx345 Cartridge

The Rx345 covers wavelengths between 0.8 to 1.1 mm. This receiver is an ALMA production type Band-7 receiver procured from IRAM. It has a quartz vacuum window explicitly designed for this frequency band. The incoming signal passes through a similar configuration as the Rx230. The signal is refocused at the off-axis elliptical cold optic mirrors and then separated into two linear polarizations with a diplexing wire grid. Each polarization signal chain consists of a sideband-separating SIS mixer. The down-converted signal at each sideband was amplified with cryogenic low noise 4-8 GHz IF amplifier, followed by a room temperature IF amplifier to reach the desired power level at the backend signal processor. This Rx345 is fully compliant with the ALMA specifications and capable of supporting the sub-mm VLBI observation. This cartridge puts out four IFs; the USB and LSB signals from two polarization channels.

### 3.5 Calibration and Test Tones

On top of the cryostat is a set of rotating arms for receiver calibration and phase stability measurement. One of the arms is attached with a piece of Eccosorb[11] large enough to block off the largest beam of the receivers. It serves as the ambient temperature load for receiver noise measurement. Attached to each of the other three arms is a phase-stable, test-tone emitter. When in operation, the designated emitter will rotate into its designated receiver signal path and transmit a tone signal to the receiver. The tone frequency is chosen to give an approximate IF frequency of around 6 GHz (see Table 5 for the exact frequencies). The tone is subsequently down-converted to 50 MHz (Final in Table 5) and sent to a vector voltmeter for phase and amplitude stability measurements against the original tone source. Confirming the receiver stability before and after VLBI observations ensures GLT performance's success during VLBI observation. More about the test tones generation will be discussed in Section 5.3.

---

[10] Millitech, model number: MXP-10-RSSSL.

[11] https://www.laird.com/

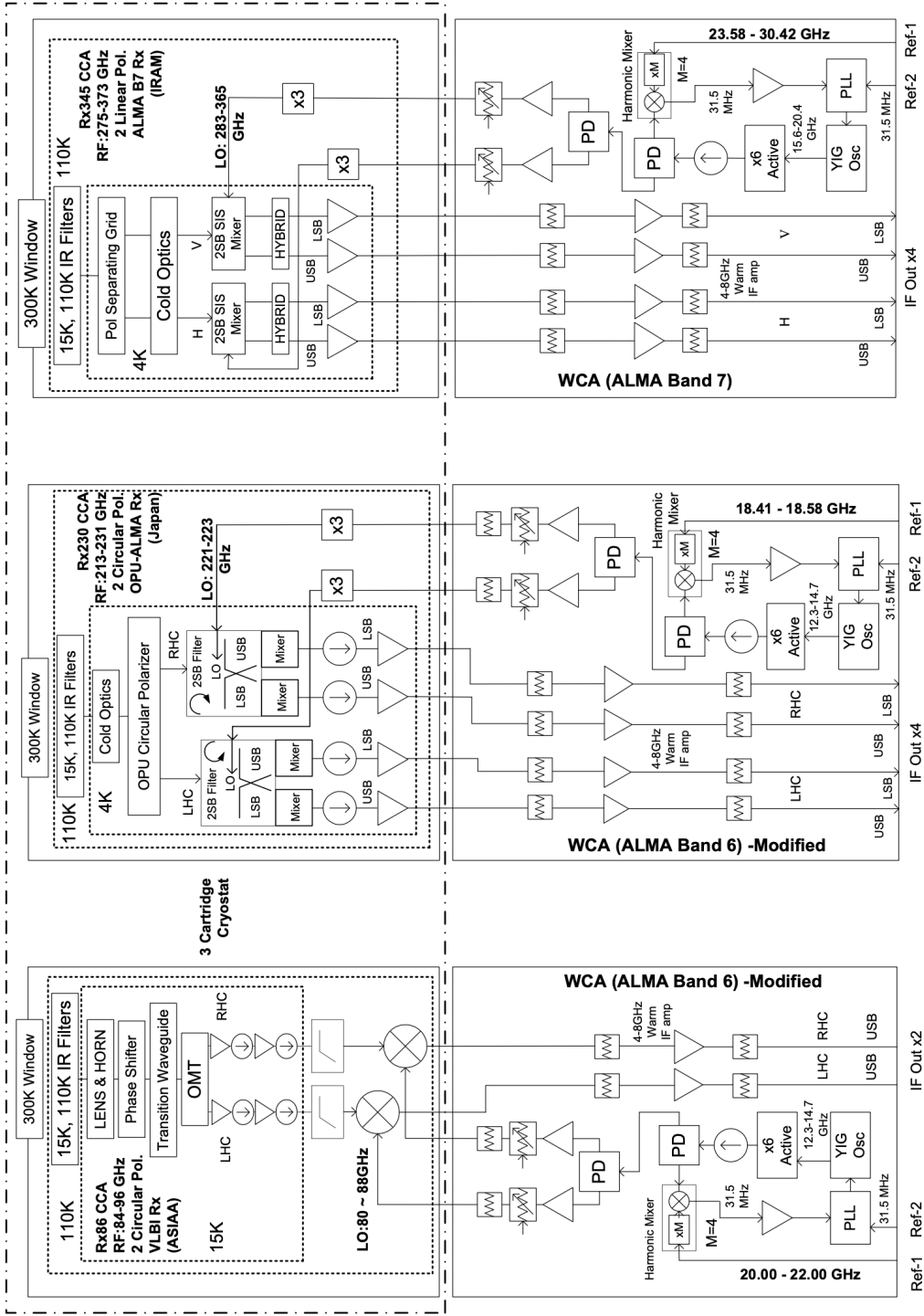

Figure 4. Simplified functional block diagram of the GLT first-light receiver system. The dot-dashed rectangle represents the cryostat that houses the three CCAs, while the WCAs are mounted with the CCAs at the bottom of the cryostat. From the top, signal from sky passes through the 300 K vacuum window, a set of infrared filters at 110 K and 15 K, and detected by the receiver front-ends. Each WCA takes in two reference signals, *Ref1* and *Ref2*, and synthesize the required LO through a unique set of phase-locked loop (PLL), Yttrium-Iron-Garnet oscillator (YIG Osc), 6-time frequency multiplier, and power divider (PD). The sky signals are down converted by the CCAs and further amplified in the WCAs before sent to the IF processors. Symbol explanation: Triangle symbols → Amplifier. Circled arrows → Isolators. Squared sawtooth → Resistor pads.

### 3.6 GLT Receiver Test in JCMT

In the summer of 2017, the GLT receiver and its associated backend systems were brought to Hawaii and installed in the James Clerk Maxwell Telescope (JCMT) in Hawaii. Being an operational telescope, JCMT provided a rare opportunity for us to test the new instrument. We recognized that much would be learned from on-sky testing that could not be easily reproduced in the laboratory. JCMT is an accessible site with local ASIAA staff who could support the commissioning. Moreover, the environmental conditions at the Maunakea summit provide a good stress test for the ancillary equipment and electronics.

The GLT instrument was successfully integrated into the JCMT system. The cryogenics of the GLT cryostat worked well at JCMT, and the GM cold head had no problems dealing with the telescope's movements. Most of the on-sky commissioning was done with the Rx230, culminating with a successful VLBI test with SMA. We also did on-sky commissioning at 86 GHz, and developed a pointing model for the Rx86 receiver. By the end of the commissioning, the GLT receiver was a functional instrument for the JCMT facility. The telescope had successfully observed CO 2-1 line from various sources and detected fringes with SMA observing 3C84 using the SMA correlator. The detail of this campaign can be found in this report [26].

### 4. Signal Distribution Networks

Two primary signal sources dictate the frequency standard and phase stability of the GLT instruments. Figure 5 shows a schematic diagram illustrating the GLT signal distribution network. Each block in the figure represents a physical subsystem handling specific signal processing, such as signal generations, multiplexing, and amplifications, etc. A hydrogen maser (Maser) produces phase-stable outputs of 10 MHz, 100 MHz, and 1-PPS for all the time-critical instruments in the GLT system. Equally important is a low-phase noise synthesizer (SYNTH) that delivers a stable tone in the frequency range of 18-32 GHz, the primary LO ($Ref\,1$ henceforth) for synthesizing the first LO signals for the receivers, and the test tones for receiver phase stability monitoring. Together with stand-alone signal sources phase-locked to the maser references, we designed and built a LO generation scheme that produces a wide range of phase-stable signals with frequencies ranging from tens of MHz for phase references to a few GHz as data recording clocks and to a few hundred GHz for RF signal down-conversion. A detailed description of the design can be found in Kubo et al.[27]

### 4.1 Maser and Timing GPS

The hydrogen maser (T4 Science iMaser 3000, S/N118) is our frequency standard for VLBI observations. It provides a precision 10 MHz reference signal with typical Allan standard deviation values of $10^{-13}$ at time intervals of 1 s and $10^{-14}$ for 10 s. It is deemed sufficient for providing low coherence loss at observation frequencies up to 690 GHz. The maser is enclosed in a thermal-controlled box that primarily provides thermal stability, and the box, as indicated in Figure 5, is physically located in a separate container away from the telescope and control container to minimize magnetic fields and mechanical disturbances. There is no additional measure for magnetically shielding for the maser other than the Corten steel of the container.

The Maser House is insulated with an active heating system to maintain stable temperatures during the critical fall through spring months. In addition, the maser itself is housed within an insulated metallic enclosure with an actively controlled heater/cooler. Any excess heat is captured by a heat exchanger and exhausted outside using a glycol loop.

We utilize a GPS unit, represented as the GPS block in Figure 5, to provide the telescope's NTP (UTC timestamp over the network), 1-PPS, and IRIG-B timing. This unit is co-located with the hydrogen maser in the Maser House container. If left freely running, the maser's 10 MHz signal might slip a few cycles per day. Hence, the maser has a 1-PPS synch input to align itself with the GPS 1-PPS. In addition, we have set up a frequency counter (Time Interval Counter in Figure 5) to monitor the 1-PPS timing difference between the GPS and maser over several days. This information is used to ascertain the frequency offset of the 10 MHz, which can be corrected via a change in the direct digital synthesizer parameter within the maser unit.

The maser's 10 MHz signal is verified regularly, as we measured its Allan variance performance before and after a VLBI observation to achieve a signal-to-noise ratio of 90 dB. It is used for phase-locking the $Ref\,1$ and in the WCAs as shown in Figure 5.

### 4.2 Fundamental LO Generation

In the Maser House, $Ref\,1$ is combined with the Maser's 100 MHz signal in the Photonic Transmission Assembly (Photonic Tx Asm) and transmitted via optical fiber to the Photonic Receiver Assembly (Photonic Rx Asm) in the receiver cabin. We used an erbium-doped fiber amplifier to compensate for the losses incurred during the transmission. Two dense wave division multiplexer devices are located on both ends of the optical fiber to combine and separate wavelengths for round-trip phase monitoring, which will be described in Section 5.2. In the Photonic Rx Asm, the optical signal is converted into radio waves and filtered to separate the 100 MHz tone from $Ref\,1$. A 10 MHz crystal oscillator is phase-locked to the maser 100 MHz and serves as the master reference for various sub-assemblies within the antenna receiver cabin. We have opted to reconstruct the master 10 MHz reference in the receiver cabin because the optical transmitter and receiver are not specified to transmit at 10 MHz. The 10 MHz phase reference is required for the offset LO generator (Section 4.3), the secondary LO (Section 4.4), the digital signal processing (Section 4.5), and the test tone generation (Section 5.3). The reconstructed 10 MHz in the receiver cabin shows a slightly degraded performance. The measurement of its phase noise yields a factor of three to four larger than that of the maser's 10 MHz signal. This degradation was negligible and did not affect the GLT's overall performance.

## 4.3 First LO and Intermediate Frequency (IF)

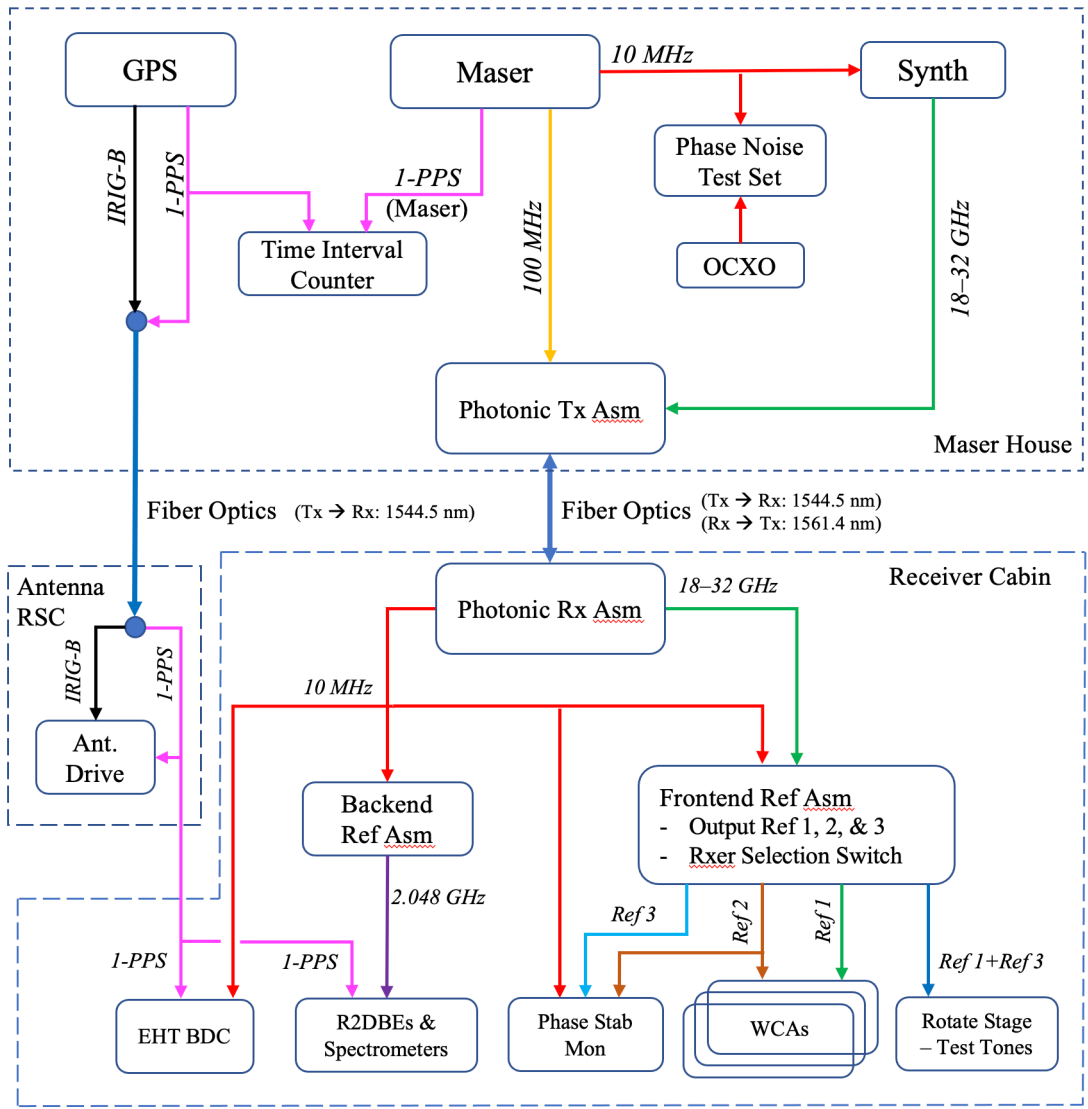

Figure 5. A schematic diagram of the GLT signal distribution network. Located in the Maser House, the maser and a noise synthesizer (SYNTH) dictate the frequency standard and stability of the GLT system. The maser's 10 MHz is monitored by a phase noise test set against a separate, free-running quartz oscillator (OCXO), while 1 PPS signal is constantly monitored by a time interval counter with an external GPS. The 100 MHz reference from the maser and the primary LO (Ref 1), in the range of 18-32 GHz, are combined in the photonic transmitter assembly (Photonic Tx Asm) and sent, via fiber optics, to the photonic receiver assembly (Photonic Rx Asm) in the receiver cabin on the antenna. In the cabin, both signals are separated and the former is used to reconstruct the master reference of 10 MHz, which is distributed to various sub-assemblies, including the frontend reference assembly (Frontend Ref Asm), phase stability monitor (Phase Stab Mon), EHT block downconverter (EHT BDC). Meanwhile, Ref 1 is sent to the Frontend Ref Asm for synthesizing the 1st LOs and the test tones of the receiver system. The 1 PPS and IRIG-B from the GPS are combined and sent, through a separate fiber, to the right-side container of the antenna (Antenna RSC). From there, the signals are distributed to various locations on the antenna for time synchronization. See text for the functions of the various assemblies.

Figure 4 depicts the receiver as a single cryostat that houses three separate cartridges: The Rx86, Rx230, and Rx345 receivers. Each receiver cartridge consisted of a CCA and its matching WCA. The LOs for the receiver front-end operations are synthesized in the WCAs and fed into the CCAs. The YIG oscillator in each WCA is chosen specifically for the matching band. Its output passes through a sixfold frequency multiplier, a phase-locked network, and a power divider to produce two identical LO signals. For the Rx86, the pair of signals are delivered directly to the mixers to produce the USB IFs of Pol-0 and Pol-1. For the Rx230 and Rx345, the LOs are up-converted for an additional three-fold in frequency before reaching the mixers. After the mixers, they produce two IFs, USB and LSB, for both polarization channels. The large frequency multiplications require the YIG signal to be high quality with low-phase noise performance.

As shown in Figure 4 and Figure 5, the Frontend Reference Assembly sends out two reference signals to set the output frequency and maintain the phase stability of the WCA. $Ref1$ is the 18-32 GHz signal from the low-noise synthesizer in the Maser House, and $Ref2$, a tone of 31.5 MHz, is generated in the Assembly. With the following formula, the LO frequency for the receiver front-ends, or the first LO frequency ($LO_{1st}$), is determined by a linear combination of these two frequencies.

$$LO_{1st} = N * (4 * Ref1 + Ref2) , \qquad (1)$$

where

N=1 for Rx86;
N=3 for Rx230 and Rx345.

The WCA needs $Ref1$ and $Ref2$ to generate its final LO. For details, please see Bryerton *et al* [23]. In the GLT system, $Ref2$ originates from a signal generator at 31.5 MHz, phase-locked to the maser reference. Both $Ref1$ and $Ref2$ are distributed to the selected WCA through a pair of switches in the Frontend Reference Assembly, as shown in Figure 5.

For VLBI observations, we do not need the Doppler tracking. When needed, it can be implemented by programming the primary LO or replacing the $Ref2$ with a programmable source.

### *4.4 IF Processing*

After the WCAs, the IF signals are sent to a pair of IF processors. Each processor can take in two IFs, and we arrange it to handle the same sideband signals from both polarization channels. The processor comprises an EHT Block Down Converter (BDC) [28] and an GLT Extension Module (EM). The BDC is a module used by the EHT partner stations and provides a unified interface during an EHT observation campaign. Since the GLT has three receiver cartridges, we developed the EM to provide the receiver selection function to the BDC input. On the output side of the BDC, the extension module also provides a replication function and delivers an identical pair of every BDC's output. The extension module is unique to the GLT among the EHT stations.

Figure 6 is a block diagram illustrating one channel of the BDC+EM. The BDC divides the IF into two "frequency blocks" and then down-converts both blocks to the baseband of 0.02-2.0 GHz. For the GLT, the IF of 4-8 GHz is divided into 4-6 GHz and 6-8 GHz and down-converted to the baseband using 6.0 GHz as the LO, generated from a signal source in the BDC. The 6.0 GHz LO is synchronized with the master reference as shown in Figure 5. However, some EHT stations use the ALMA Band-6 receiver with an IF range of 5-9 GHz. In such a case, the BDC is equipped with separate paths that divide the IFs into 5-7 GHz and 7-9 GHz and use 7.0 GHz as the LO for down-conversion. The BCD achieves the band separation using high rejection band-passed filters.

In the case of the GLT, only one of the three VLBI receivers is active during observation. The receiver (except the Rx86) produces four IFs. The BDC processes each IF input into four baseband signals, filtered to meet the Nyquist sampling criteria and level-controlled using variable attenuators to satisfy the drive levels into the digitizers. The pair of BDC generate eight baseband blocks representing:

- POL-0 LSB 4-6 GHz
- POL-0 LSB 6-8 GHz
- POL-0 USB 4-6 GHz
- POL-0 USB 6-8 GHz
- POL-1 LSB 4-6 GHz
- POL-1 LSB 6-8 GHz
- POL-1 USB 4-6 GHz
- POL-1 USB 6-8 GHz

In the case of the Rx86, only the USB blocks are available. All the basebands generated from the BDC are replicated in the extension module, producing two copies of each baseband for the following digitization process.

All of the controls for the switches, amplifier gains, and variable attenuation are remotely controllable over the network.

### *4.5 Baseband Processing*

The IF processor generates two copies of each baseband signal. One copy is sent to a field-programmable gate array (FPGA) module that digitizes and packetizes the baseband signal in the format for VLBI data recording. The other copy is sent to a similar FPGA module of a different function, programmed as an auto-correlation spectrometer.

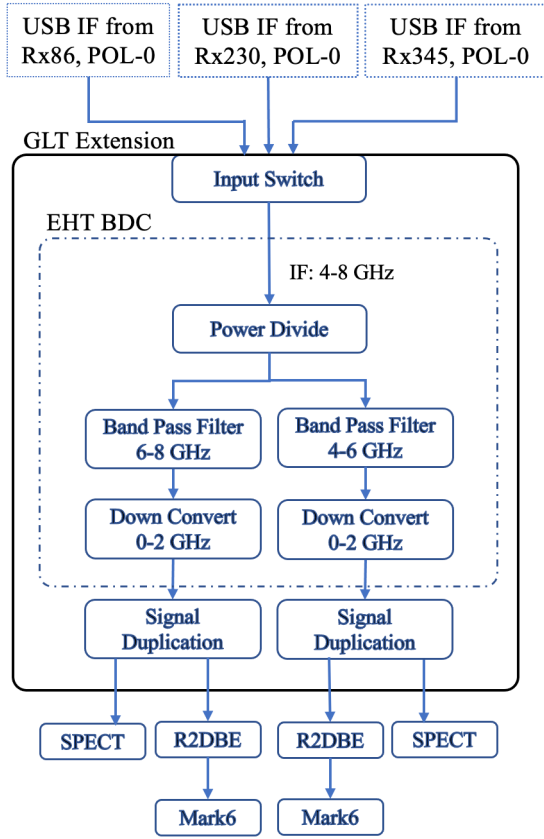

Figure 6. A block diagram illustrating the signal processing of the EHT Block Down Converter and the GLT Extension Module (EM). As shown, the GLT EM selects the receiver input and passes one IF to the BDC, which then converts the IF into two blocks of baseband signals, each with 0-2 GHz bandwidth. The EM duplicates the blocks and output an identical pair of each baseband. The BDC+EM module can process two IFs signals at the same time.

The FPGA module is built on the ROACH-2 [12] unit developed by the CASPER[13] collaboration. A ROACH-2 unit comprises a main board with a Xilinx Virtex-6 FPGA with two 5 Giga-Sample-per-second (GSps), 8-bit analog-to-digital converter (ADC) [29] as the input. For the VLBI application, the EHT consortium built the ROACH-2 into a standard digitizing backend for the VLBI recording. The R2DBE (ROACH-2 Digital Band End) [30] [31] takes in two baseband signals. With a clock rate of 2.048 GHz, each R2DBE sampled at a rate of 4.096 GSps and 8-bit quantization to produce 65 Gb of data per second. Transmitting such a large amount of data with commodity hardware is impractical. But for astronomical VLBI applications, a time stream of 2-bit data is sufficient for correlation, albeit with some small loss in sensitivity. The 8-bit data is thus truncated to 2-bit so that the outgoing data rate is reduced to 8.192 Gbps, with overhead excluded. After formatting, the R2DBE transmits the repackaged data to the Mark-6 VLBI data recorders[14] through two 10-Gigabit Ethernet (GbE). Each Mark-6 can record two basebands at the rate of 16 Gbps. The GLT has four sets of Mark-6 running simultaneously to record 64 Gbps of data.

The auto-correlation spectrometer is built on the same ROACH-2 hardware with two ADC inputs. Like the R2DBE, each ADC takes one baseband signal at a rate of 4.096 GSps and 8-bit quantization. The ROACH-2 combines two ADC inputs, performs digital filtering (polyphase filter bank), and transforms the data into a spectrum of 4.096 GHz bandwidth with 65,536 channels, each represented by a 4-bit + 4-bit complex number. The spectral data is accumulated in the FPGA for a specified integration time and then packetized for sending over the Ethernet. The spectrometer can generate a spectral frame every few micro-seconds. However, this application does not need such a high data rate. The resultant spectral information is sent to a spectral monitoring computer over a 10 GbE cable.

During observations, one of the receivers puts out four IFs (the Rx86 only outputs two IFs). After the IF processors, the IFs are turned into two identical sets of eight baseband blocks. One set of blocks is sent to the R2DBE, while the other is to the spectrometers. We only have two spectrometers to handle the USB signal and will install another two for the LSB.

## 5. Calibration and Diagnostic Test System

Since the GLT is located at a remote site, it takes considerable time to transport the VLBI data to the correlator centers. Any problems encountered during the VLBI observations are only detected much later. Hence, we implemented several calibrations and diagnostic tests to ensure the proper working of the entire GLT system before and after the VLBI observations. We are primarily concerned with the stability of the frequency standard of the hydrogen maser, the integrity of the LO distribution path between the maser and the antenna, the stability and sensitivity of the receiver and the subsequent signal processing system, and the working condition of the Mark 6 recorders. The following sections describe the tests and the methods.

---

[12] ROACH-2 is a stand-alone FPGA board and is the successor to the original ROACH board. ROACH stands for Reconfigurable Open Structure Computing Hardware. See <https://github.com/casper-astro/casper-hardware#casper-hardware>, as of Dec 2022.

[13] CASPER stands for The Collaboration for Astronomy Signal Processing and Electronics Research.

[14] MIT Haystack, "The mark 6 system has been developed by the Haystack Observatory as a 16 Gbps next generation disk-based VLBI data system based on high-performance cots hardware and open-source software," <https://www.haystack.mit.edu/tech/vlbi/mark6/index.html>.

*5.1 10 MHz Maser Performance Monitor*

One of the most critical parameters for millimeter-wave VLBI operations is the purity and stability of the master reference, the 10-MHz signal from the hydrogen maser. Generating an LO of 221.0 GHz [28] requires scaling the master signal by a factor of 22,100, and minute jittering of the reference will induce significantly undesirable effects on the LO. The maser unit has a web-based telemetry function that alerts the user if its internal parameters are outside specification tolerances. For additional insurance, we use an external phase noise test set (Symmetricom TSC 5120A) and compare the maser's 10 MHz against a separate quartz oscillator (Oscilloquartz model 8607). The test generates the Allan standard deviation plots. Because the stability performance of the maser is about two orders of magnitude better than that of the quartz oscillator, the measurements should only reach the stability limit of the quartz oscillator, not the maser. Nevertheless, this setup provides a secondary verification of the maser's performance within our technical capabilities and financial restrictions.

*5.2 Round Trip LO Phase Monitor*

As previously mentioned, the hydrogen maser is located separately from the rest of the GLT system. The maser signals are delivered to the receiver cabin through optical fibers. In expecting that the fibers might run for a few kilometers on the Greenland summit, we have devised a round-trip, phase measuring method to monitor the signal stability of the transmission. The setup was built into the Photonic Transmitter Assembly (shown in Figure 5) in which it combines a 15.75 GHz pilot tone with the 18-31 GHz LO and 100 MHz maser reference before optical transmission at a wavelength of 1544.5 nm. Received in the Photonic Receiver Assembly, the pilot tone is separated from the two other signals using a band-pass filter. It is optically re-modulated at a wavelength of 1561.4 nm and sent back to the Maser House. The returned tone is separated from the optical carrier and compared with the original pilot tone. The comparison takes place in a vector voltmeter after both tones are down-converted to 10 MHz using an LO at 15.76 GHz in the Photonic Transmitter Assembly. The resultant variations of amplitude and phase between the tones represent the distortions induced by temperature and mechanical stresses onto the fiber carrying the 15.75 GHz pilot tone. During a typical VLBI observation, the received signal is integrated for up to 10 seconds, while the monitored phase variation is usually less than half a degree RMS at 15.75 GHz. Scaled up to the 1st LO, e.g., 221.1 GHz, the phase error is multiplied to 5 degrees RMS. Such a phase error will only affect the sensitivity by a few percent [32] and is minor compared to those caused by atmospheric effects at higher sub-mm frequencies. However, the situation might not be such trivial when the LO must travel a long distance to reach the antenna. The round-trip phase monitor provides a means for post observation phase correction. In the meantime, we are investigating methods for actively compensating for changes in this round-trip phase.

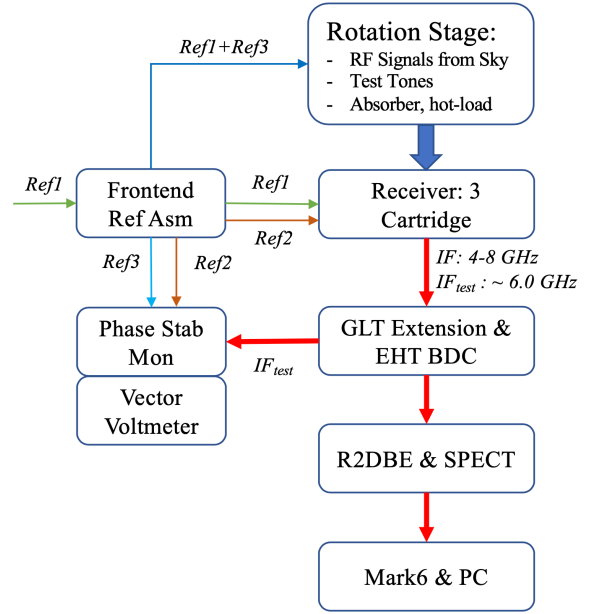

Figure 7. A block diagram depicting the generation and flow of the test-tone in the GLT receiver system.

*5.3 Test Tones and Reference Signals Generation*

Verifying the receiver stability before and after VLBI observations is important for GLT operations. As described in Section 3.5, each of the three rotating arms carries a specific tone emitter composed of a harmonic mixer and a feedhorn to generate a linear-polarized test tone for its designated receiver band. The harmonic mixers up-convert the incoming LOs and output x4 (for Rx86) and x12 (for Rx230/Rx345) tone signals. Referring to Figure 5, a dual-output synthesizer in the Frontend Reference Assembly generates two stable signals of 500 MHz and 1.5 GHz. Depending on which receiver band is chosen, one of the signals (referred to as $Ref3$ below) is selected to combine with $Ref1$ (see Section 4.3) and generate a tone of summed frequency. The frequency of the test tone, $RF_{tone}$, will be the multiplication of the sum and can be calculated from the following formula:

$$RF_{tone} = M * 4 * (Ref1 + Ref3), \qquad (2)$$

where:

M = 1 and $Ref3$ = 1.5 GHz for Rx86;

Table 5
The Summary of the Test Tone Signal Flow

| Keyword | Rx86 | Rx230 | Rx345 |
|---|---|---|---|
| $N^i$ | 1 | 3 | 3 |
| $M^i$ | 1 | 3 | 3 |
| $Ref\,1^{ii}$ | 21.492125 | 18.417125 | 26.742125 |
| $Ref\,2$ | 0.0315 | 0.0315 | 0.0315 |
| $Ref\,3$ | 1.5000 | 0.5000 | 0.5000 |
| $RF_{tone}$ | 91.9685 | 227.0055 | 326.9055 |
| $LO_{1st}\,(\downarrow)^{iii}$ | 86.0000 | 221.1000 | 321.0000 |
| $IF_{1st}$ | 5.9685 | 5.9055 | |
| $LO_{2nd}\,(\downarrow)$ | 5.9500 | 5.9500 | |
| $IF_{2nd}$ | -0.0185 | 0.0445 | |
| $LO_{final}\,(\uparrow)^{iv}$ | +Ref 2 | +3 * Ref 2 | |
| Final | 0.0500 | 0.0500 | |

$^i$ : N and M are defined in Equations (*1*) and (*2*).
$^{ii}$ : All frequencies are in GHz.
$^{iii}$ : The down arrow indicates a down-conversion process.
$^{iv}$ : The up arrow indicates a up-conversion process.

M = 3 and $Ref\,3$ = 500 MHz for Rx230 and Rx345.

The test tone is sent toward its designated receiver channel. After the first down-conversion in the receiver, the resultant IF signal ($IF_{1st}$) is tapped out from the GLT Extension Module (before entering the EHT-BDC) and sent to a phase stability monitor (Phase Stab Mon, as shown in Figure 5). In the phase stability monitor, $IF_{1st}$ is mixed with a secondary LO of 5.950 GHz ($LO_{2nd}$) to generate a secondary IF ($IF_{2nd}$). By design, the resultant tone signals from the mixer output are 18.5 MHz for the Rx86 and 'negative' 44.5 MHz for the Rx230 and Rx345. The negativity indicates that the signal's phase rotates oppositely to that of the master reference. The 44.5 MHz tone is then up-converted with a reference signal ($LO_{final}$) at 94.5 MHz, generated from the three-time multiplication of $Ref\,2$, while the 18.5 MHz tone is summed directly with $Ref\,2$. Finally, the test tones from the three receiver channels are all down-converted to a 50 MHz tone signal, which is band-passed and sent to the input port of the vector voltmeter.

Not only the frequencies of $RF_{tone}$ and associated $LO_{1st}$ are selected to be near that of the VLBI observations, but they also must fit into the multiple down-conversion schemes described above. Figure 7 provides a block diagram of the test tone signal flow in the receiver system. The scheme for generating the test tone signals is tabulated in Table 5.

The reference input of the vector voltmeter is generated from the multiplications of the $Ref\,3$ signal. The $Ref\,3$ signal, either 500 MHz or 1.5 GHz, is multiplied by 12 or 4, respectively, to produce a desired IF of 6.000 GHz and is then subsequently mixed with to produce the 50 MHz tone to the reference port of vector voltmeter.

The graph presented in Figure 8 represents the long-term phase (difference) and amplitude (ratio) stability of the Rx230 over eight hours starting at 11:30 pm on April 21, 2018. The phase gradually increased by approximately 50 degrees over four hours (12.5 degrees per hour), then flattened out for the remaining four hours. A 50-degree phase movement at a test tone frequency of 227.0055 GHz corresponds to a physical movement of 0.184-mm between the tone source on the swing arm above the receiver and the receiver itself. This phase drift is smaller than the phase error discussed in Section 5.2 and can be corrected during calibration and post-processing.

### 5.4 Fiber Optics System

All RF reference signals and data communications to and from the telescope are transmitted over single-mode fiber optic cables capable of supporting 1310-nm and 1550-nm optical windows. Designed to withstand temperatures down to -65°C, the cable consists of a 1.32-mm OD 316 stainless steel tube that contains four polyimide-coated fibers. The stainless-steel tube is spirally wrapped by #12 316 stainless steel wires and a 3.80-mm OD HDPE outer jacket. We selected this small diameter 4-fiber cable because of the small dynamic bend radius of 132 mm (5.2 inches). The GLT system consists of twelve fiber optic cables routed from the Control trailer, VLBI trailer, and the Maser House to three locations within the telescope. Note that all twelve fiber cables must pass through the telescope azimuth wrap, with eight extended further to pass through the elevation wrap and enter the receiver cabin. The current system utilizes seven of the twelve cables.

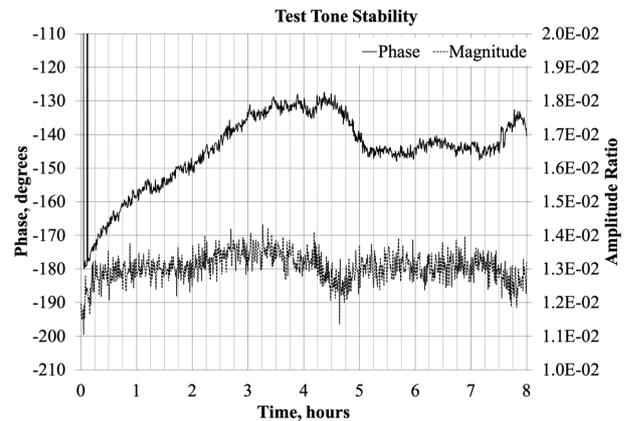

Figure 8. The Rx230 test tone phase and amplitude response. A 227.0055 GHz test tone was generated and transmitted to the receiver, where it was mixed with a first LO of 221.1000 GHz to produce an IF of 5.9055 GHz. This IF tone was mixed and converted to 50 MHz within the continuum detector and phase stability monitor units, then the phases and amplitudes are compared to a 50 MHz reference.

## 6. Dish Surface Accuracy

As mentioned in Section 2.3.1, the GLT primary dish, when first put together on the ground, was aligned and set to an accuracy of 40 μm RMS using a laser tracking technique. The accuracy degraded after the dish was installed onto the antenna but then was recovered with panel adjustments based on the results from photogrammetry surveys. We achieved a surface accuracy of 17 μm RMS as the best result.

However, the antenna's surface accuracy degrades as the ambient temperature changes from the time of the panel setting. Hence, one should survey the dish surface and optimize the accuracy on a dark winter day, during which no solar heating occurs. The atmosphere offers the best atmospheric transparency for observation in sub-mm wavelengths. In principle, we can carry out a photogrammetry survey in winter. We are also developing a near-field holography method, described in a later section, to survey the primary dish in winter. But we face a safety issue: the dish panels can only be manually adjusted, and a complete adjustment of the primary dish is an outdoor task that lasts several hours. This is not a practical exercise during the Arctic winter. We are currently exploring the simulation and modeling techniques to resolve this issue.

### *6.1 Photogrammetry*

Photogrammetry is a 3-dimensional coordinate measuring technique that uses photographs as the primary medium for metrology. It uses triangulation to acquire, from different angles, at least two photographs of every object on the surface of interest to reconstruct the 3-dimensional model. These objects are the small reflective targets on the dish surface beside the panel adjusters. In addition, a few coded objects are recognizable by the processing software for determining the camera position and orientation when a picture is taken. The data are taken using a specially modified camera[15] with flash lighting, which obtains hundreds of strobed photographs taken at various viewing angles from the camera to the primary. The photographs are the digital images of the reflective targets, the coded targets, and the two reference points. They are processed by proprietary software to derive the relative positions of each target, which are then fitted with the theoretical paraboloid to determine the surface accuracy.

In 2012, we conducted two photogrammetry surveys on the Prototype at the VLA site; one was carried out at an antenna elevation of 5°, and the other at 90°, and the results of the surface accuracy were 78 μm RMS and 79 μm RMS, respectively. Both results showed a similar surface pattern of astigmatism. It is probably due to the high stiffness of the antenna that the surface accuracy does not vary with the antenna elevation angle. Later, after a few iterations of panel adjustments and surface surveys, we set the dish surface to an accuracy of 40 μm.

After the GLT was assembled in Pituffik, the first photogrammetry in August 2018 resulted in a surface accuracy of 170 μm RMS, even though the dish was aligned to 40 μm

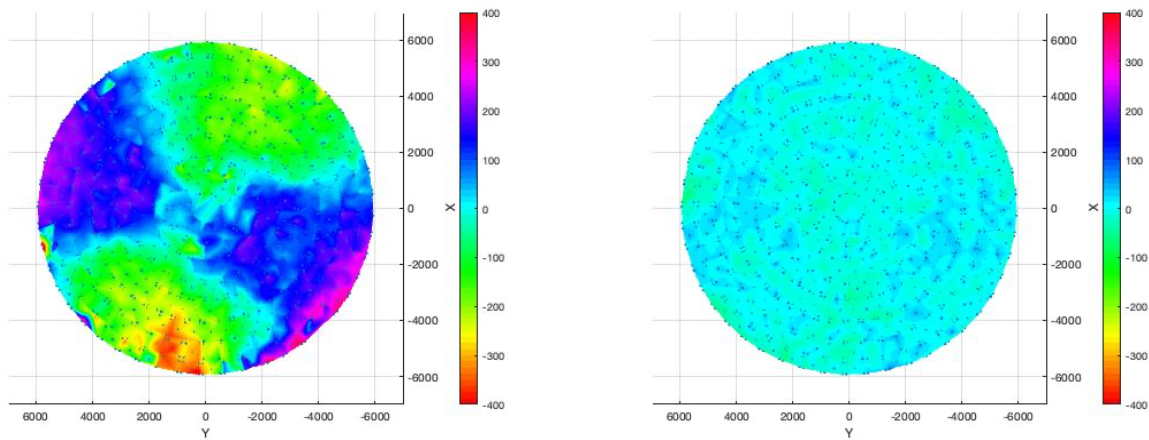

Figure 9. Surface error maps taken with the photogrammetry measurements. On the left is the map representing a surface accuracy of 170 μm RMS. It shows a saddle pattern consistent with the locations of the attached harness for lifting the dish onto the telescope mount. On the right is the map resulting from several runs of photogrammetry and panel adjustments. The result, 17 μm RMS, is the best surface accuracy achieved since the completion of the GLT in 2018. The colored bars and the coordinates are in units of μm.

---

[15] V-Stars system from Geodetic Systems, Inc. (GSI)

RMS before the assembly. The surface was eventually set to 17 μm RMS. Figure 9 shows the surface error maps of the two extremes. The final panel adjustment process was done at the ambient temperature between 2-5°C.

The surface accuracy degraded when the ambient temperature changed. The degradation was verified in the aperture efficiency measurements during the Pituffik winters (~ -30°C), and the dish efficiency implied a surface error of around 40 μm RMS. In April 2019, at the ambient temperature of -15°C, we measured the surface again with photogrammetry and found a surface error of 26 μm RMS.

For the Prototype, Mangum et al [5] reported that the magnitude of the surface RMS deformation is about 0.6-0.7 μm /K. According to VA's FEA modeling, about 1/3 of the deform can be attributed to the panels and the rest to the BUS. Assuming all effects are linear with temperature, the change of surface RMS from 17 μm, which was set near 0°C, to 40 μm at -30°C, is consistent with the previous experience.

The precision of the photogrammetry method is affected by the camera, size of the measured volume, survey technique, ambient conditions, and target reflecting light quality, etc. The measured results should include all the errors from these factors. GSI told us that the precision of photogrammetry measurement on a 12-meter dish should be around 45 um RMS. However, we constantly achieved values better than 20um RMS. The fact that our best results were achieved through an iterative process of surface adjustments followed by repetitive measurements convinced us they should be genuine and valid.

### *6.2 Holography*

As an alternative for surveying the dish surface accuracy, we are developing a near-field holography system for the GLT, following the experience of the SMA and APEX telescope [33, 34, 35]. The holography beacon is mounted on a tower atop the ridge toward the south of the telescope. It emits a tone at 94.5 GHz toward the GLT and is received by the Rx86. In the meantime, the tone is also received by a reference receiver mounted on the back side of the GLT sub-reflector. The tower's location is at the GLT's elevation angle of 2.9° and is about 2.9 km away from the antenna. Because the GLT has a mechanical lower limit of 1.9° in elevation, the maximum map size is limited to about 2 deg$^2$ in size. We only covered a square map of 1.46° x 1.46°.

With on-the-fly mapping of the beacon, the GLT carries out a series of azimuth scans at a constant rate of 0.075°/s. The amplitude and phase measurements are made using a vector voltmeter read out along with the antenna encoders once every 50 ms. The data are re-gridded on a 128 x 128-pixel map and inverted to obtain the illumination pattern and the surface error maps across the antenna aperture. With this map, we can obtain 14.0 cm resolution on the antenna dish surface.

There is a power unbalanced issue due to the significant difference in receiver gains between the Rx86 and the reference receivers. Under consideration is the idea of implementing a dedicated holography receiver to replace the Rx86 in the receiver cabin.

## 7. Control and Communication

The GLT control and monitoring system comprises over a dozen subsystems linked via Ethernet and time synchronization networks. Each of the subsystems has its controller with a unique operating system. The software is a collection of Linux and Windows-based programs. The observers interact with the GLT control and monitoring system through a common console, dubbed the Observation Console. The following sections describe the inter-working in the GLT control and monitoring system.

### *7.1 Antenna Drive System*

Figure 10 shows a block diagram of the GLT control system. The GLT antenna drive system is part of the retrofitting contract with VA. Located in one of the antenna side containers, the Antenna Control Unit (ACU) coordinates all actions within the servo system and interfaces to external equipment. It also accepts a limited set of parameters for pointing correction. The operator's interface at the ACU is a color touchscreen panel. We can also access the ACU's full functions remotely via an Ethernet TCP/IP interface. The azimuth and elevation axes have anti-backlash drive units with brushless motors. A set of servo amplifiers provide the motors' currents. Servo amplifiers, as well as the entire control loop implementation, are fully digital. A local bus system connects amplifiers and ACU.

The antenna has an interlock system comprising a programmable logic controller (PLC) overseeing any inadmissible state. The interlock includes absolute optical encoders measuring actual positions, limit switches for mechanical motions, and other safety functions. The PLC reads various metrology sensors and forwards the readouts to the ACU. Some sensors can be used for pointing corrections, others for monitoring purposes only. The interlock does not sense extraordinary accelerations or oscillations.

From the ACU, we can command the antenna for movements such as pointing the antenna to a fixed azimuth (AZ) and elevation (EL) position, stowing the antenna, or holding either AZ/EL position while scanning at a fixed rate on one or both axes, etc. The ACU can also issue tracking commands to observe astronomical sources, though with less precision than required for the GLT science operations. It can also operate in the Two-Line mode to track artificial satellites.

Typically, the telescope is operated "remotely", or off the antenna, from a dedicated station computer (STC, referred to as the Antenna Control Computer, ACC, in the GLT system) located in a nearby trailer. The ACC, ACU, and the rest of the computers involved in the real-time operation are time-synchronized via IRIG-B signal from a GPS unit. The GLT users access the ACC from the Observation Console via a dedicated Ethernet connection. The ACC provides a platform for implementing user-defined control and monitoring software for various needs of astronomical observations. The software is based on codes developed for the Submillimeter Array (SMA). A detailed account of the process workflow can be found in Patel et al [36, 37, 38].

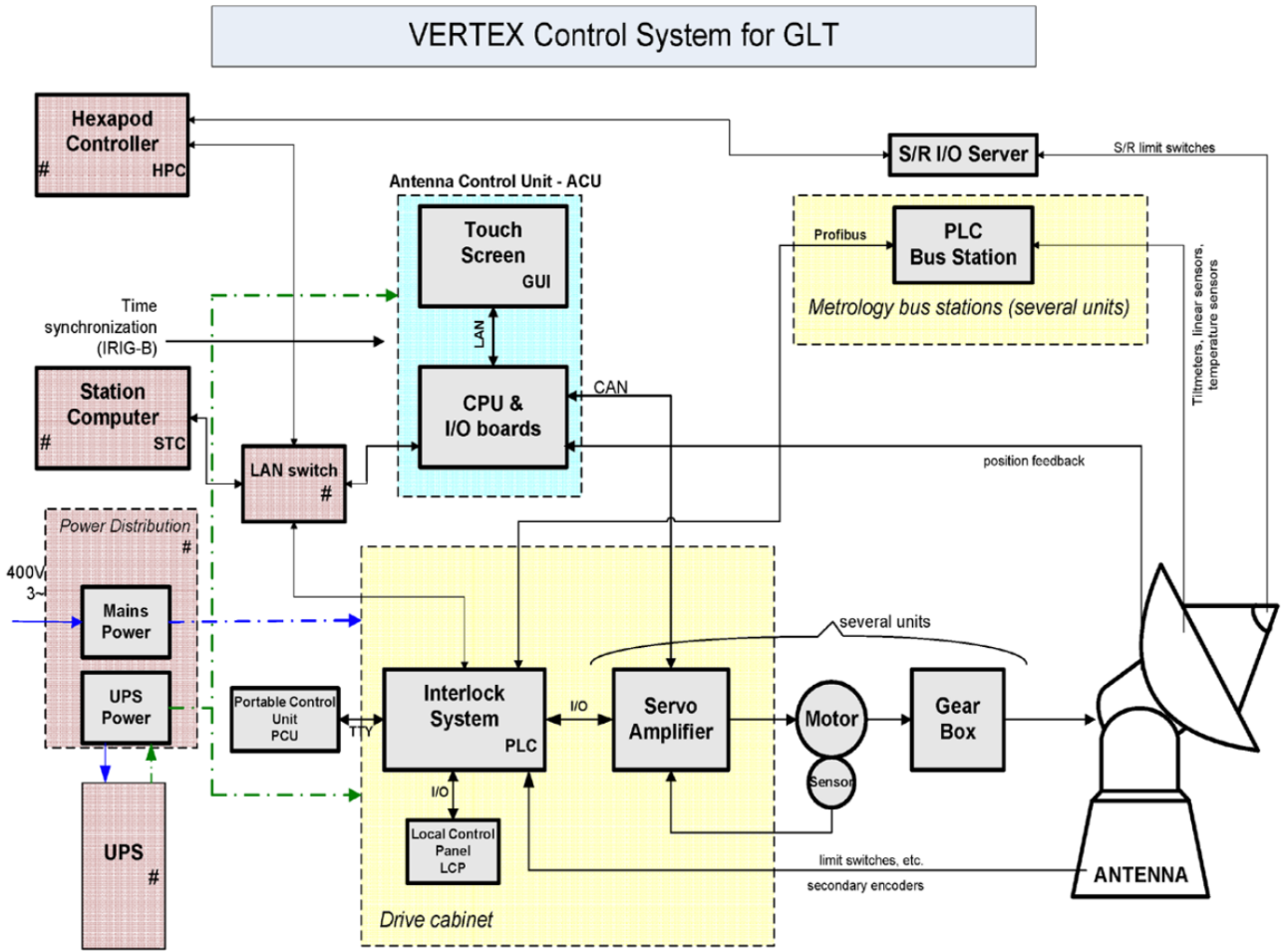

Figure 10. Block diagram of the GLT antenna control system. The antenna motion control is driven by the Antenna Control Unit (ACU), which permits manual control via the touchscreen panel and provides an ethernet interface for control and monitoring. The Antenna Control Computer provides higher-level control, accepting user commands and communicating with the ACU. The Hexapod Control Unit provides similar functionality as the ACU but for the sub-reflector hexapod position control. All other boxes in this diagram are part of the VA antenna drive system components, enclosed in the servo cabinets in the antenna's right-side container. The blocks marked with "#" are the equipment provided by the GLT team.

### 7.2 Hexapod Drive System

The hexapod comes with its controller and is not part of VA's delivery. The Hexapod Controller provides interfaces through Ethernet for users' control and monitoring. The typical commands include positioning and tilting the sub-reflector, starting and ending the various modes of nutation, and querying the system status of the hexapod. Section 2.1.2 has described the hexapod's basic operations.

The Hexapod Controller communicates with the ACC via ethernet but does not have an IRIG-B interface for time synchronization. This has limited the hexapod's chopping rate to be less than 1 Hz because the interface refreshes at the time scale ~ 1 sec. We currently explore means of synchronization to utilize the hexapod's capability fully.

### 7.3 Receivers and Detectors

We developed a software suite for the receiver operations, e.g., setting up the synthesizer frequency, power-tuning the LO, side-band selection, controlling the rotation stage of injected tones/loads, monitoring several front-end devices, etc. We utilized the ALMA receiver software package for biasing and phase-locking the warm cartridge assemblies for the desired frequency—a more detailed description to be reported shortly.

## 8. Pointing and Efficiency

Science commissioning activities for the GLT began in December 2017. Establishing an all-sky optical pointing model was straightforward, using an optical guide scope mounted on the antenna dish. The optical pointing model was the starting point for the following radio pointing. With the Rx86, we observed the Moon, which gave us the coarse offset of the radio pointing. The Moon's edge provided convenient spots to set the initial focusing positions of the sub-reflector at 86 and 230 GHz. The foci were further fine-tuned by observing Venus once the planet was above the horizon, and, finally, the GLT was sensitive enough to observe spectral line radio sources, such as SiO masers at 86 GHz and CO(2−1) at 230 GHz, and establish the all-sky radio pointing model. With the radio pointing model ready, the GLT participated in the EHT dress rehearsal observations in January 2018, and completed the EHT science observing campaign in April 2018. Fringes were successfully detected for these early VLBI experiments for the GLT[11].

The GLT's surface accuracy was improved in the summer of 2018, and so was the antenna aperture efficiency. This has enabled the GLT to detect weaker molecular lines and sources during test observations. Through all-sky spectral line pointing observations (SiO masers at 86 GHz and CO at 230 and 345 GHz), we achieved radio pointing accuracy down to ∼ 2″ at all three receiver bands. The sub-reflector focusing was further fine-tuned following the surface accuracy improvement. We detected fringes between the GLT and other stations for most VLBI experiments. During the pandemic period (2020-2022), we have transitioned to a remote operation mode. The following section provides a brief summary of the scientific commissioning activities at the GLT, including the first light and test observations at 345 GHz. A more detailed account can be found in Koay et al [39].

### *8.1 Pointing Model*

We use the following equation to model the elevation offsets ($\Delta EL$) as a function of telescope elevation, $EL$, and azimuth, $AZ$:

$$\Delta EL = eldc + elsagcos * \cos(EL) + elsagsin * \sin(EL) + eaztltsin * \sin(AZ) + eaztltcos * \cos(AZ) + eaztltsin2 * \sin(2*AZ) + eaztltcos2 * \cos(2*AZ) \ . \quad (3)$$

Table 6
The Comparison of the pointing models between the GLT and Prototype.

| Coef. in GLT Pointing Model | Coef. in Ref.[5], for ALMA NA Prototype | Nominal Cause |
|---|---|---|
| eldc | +IE | EL encoder zero-point offset. |
| elsagcos | +HECE | Gravitational sag/Hooke's Law vertical flexture/EL encoder run-out, cosine component. |
| elsagsin | +HESE | Gravitational sag/Hooke's Law vertical flexture/EL encoder run-out, sine component. |
| eaztltsin | +AW | Telescope tilt due to azimuth axis misalignments/ Az axis offset/misalignment east-west (west = positive). |
| eaztltcos | -AN | Telescope tilt due to azimuth axis misalignments/ Az axis offset/misalignment north-south (north = positive). |
| eaztltsin2 | +HESA2 | Telescope tilt due to azimuth axis misalignments/ El nod twice per Az rev, sine component. |
| eaztltcos2 | +HECA2 | Telescope tilt due to azimuth axis misalignments/ El nod twice per Az rev, cosine component. |
| azdc | -IA | Az encoder zero-point offset. |
| azcoll | -CA | Non-perpendicularity between the boresight and the elevation axis. |
| aeltlt | -NPAE | Non-perpendicularity between the Az and El axis. |
| aztltsin | -AN | Az axis offset/misalignment north-south (north = positive). |
| aztltcos | +AW | Az axis offset/misalignment east-west (west = positive). |
| aztltsin2 | +HVSA2 | Az/El nonperp variation twice per Az rev, sine component. |
| aztltcos2 | -HVCA2 | Az/El nonperp variation twice per Az rev, cosine component. |

The azimuth offsets are modeled using the following equation:

$$\Delta AZ = azdc + azcoll * \sec(EL) + \\ aeltilt * \tan(EL) + \\ aztltsin * \tan(EL) * \sin(AZ) + \\ aaztltcos * \tan(EL) * \cos(AZ) + \\ aztltsin2 * \tan(EL) * \sin(2*AZ) + \\ aztltcos2 * \tan(EL) * \cos(2*AZ) \quad . \quad (4)$$

The pointing offset model, or the pointing model, evolved from a combination of two previous works; one is for the pointing calibration of the SMA [40], and the other is for the Prototype [5]. Table 6 provides a comparison between the GLT and the Prototype. The first and second column of the Table lists the corresponding coefficients used in both models. The third column briefly lists the nominal causes of each term.

### 8.2 Antenna Pointing

An optical telescope, SBIG ST-i, is mounted inside a hole on the primary dish (see Figure 3) and is used for optical pointing measurements. When taking pointing data, we run custom software that searches for and captures images of stars on a strip of sky, either vertically (between 20° to 80° in EL for a fixed AZ), or horizontally (between 0° to 360° in AZ for a fixed EL). For a typical optical pointing run, we make these observations at horizontal strips of EL = 20°, 40°, 60°, 80° followed by vertical strips at AZ = 45°, 90°, 135°, ... 325°, where around 100 to 200 images would be obtained depending on the sky conditions [11].

Once the run is complete, the pointing offsets are determined based on the EL and AZ offsets of the star from the center of the optical image. We derive the pointing model coefficients by fitting the same pointing model to $\Delta EL$ and $\Delta AZ$, Equations (3) and (4), as functions of telescope elevation and azimuth.

Obtaining the radio pointing takes a few additional steps. At each nominal location, we measure the target's spectral line intensity at five positions: at the expected source center and ±X arc-second in azimuth and elevation; X is typically selected to be the half-width at half maximum of the primary beam (sometimes adjusted depending on source compactness). The target's position is determined by fitting the five-point intensity data to a Gaussian distribution, and its difference with the nominal position is the pointing offset.

For the radio pointing, we observe SiO J=2−1 maser sources at 86 GHz, including o Cet, NML Tau, TX Cam, IRC+60169, R LMi, R Leo, R Cas, and χ Cyg. At 230 and 345 GHz, we observed CO J=2−1 and CO J=3−2 sources, including IRC+10216, CIT 6, NGC 7027, CRL 2688, CRL 618, IRAS 21282+5050, R Cas, and χ Cyg.

We conduct the pointing observations before each VLBI science observing season to verify the pointing accuracy and to correct for minor changes in the pointing model over time. Based on our experience, the RMS pointing accuracy remained within 3" up to several weeks after an all-sky reading from tiltmeters installed on the telescope yoke arm and support cone to monitor changes in telescope tilt relative to time and ambient temperature. The antenna tilt variation was also reflected in the changes in the radio pointing model.

After the first winter season (2018 Dec - 2019 Apr), the GLT's pointing errors at 230 GHz and 345 GHz were within the telescope specification of 2", but at 86 GHz was slightly larger due to its larger beam size.

### 8.3 Station Position

Accurate information about the station position is critical for VLBI correlation. While the position can be determined more accurately during the correlation process once fringes are detected, an initial estimate is needed first to detect these fringes. For VLBI observations at 230 GHz, an accuracy of < 5 m is needed for a fringe rate of < 100 mHz.

The station position is typically defined as the intersection of the azimuth and elevation axis of the telescope since this position is always stationary regardless of the movement of the antenna. Some VLBI stations measure this position of intersection directly relative to the reference position associated with the GPS. For the case of the GLT, we installed a GPS on the roof of the left side container. As the telescope rotates in the azimuth direction, the GPS also rotates with respect to the azimuth axis.

On the other hand, the GPS does not move when the telescope moves in the elevation axis. Therefore, we can estimate the position at the height of the GPS on the azimuth axis by measuring the GPS position at different hour angles by rotating the telescope. We conducted this measurement on 2018 Jan. 23. For this measurement, we rotated the telescope every 60 degrees and obtained GPS positions for 10 minutes at each position. After fitting to the circle and deriving the origin of the circle, we obtained the position of the station. The data thus acquired on Jan 23, 2018 was (76°32'6.5767" N, 68°41'8.8237" W), and Altitude: 89.40 m. For comparison, at Summit Station, the telescope location would be at roughly at (72°34'46" N, 38°27'32" W).

We note that an additional offset between the center of the circle and the intersection of the azimuth and elevation axes should remain. However, it should be within 1 mm by design. This scheme is adequate for the station position measurements at the Summit, as such measures can be completed within a few hours.

### 8.4 Snow and De-ice Effects

Snow accumulation on the primary dish affected the telescope pointing and aperture efficiency. On one occasion, when the snow, about a millimeter thick, covered the lower half of the dish surface, it more than doubled the elevation offset at the 230 GHz pointing exercise. We noticed a lesser effect in the azimuthal direction.

The snow on the primary dish, usually in the lower half, likely causes irregular absorption leading to a non-uniform illumination pattern. We could clear the snow with the dish turning toward the direction of the Sun. During the Arctic winter season, we used the de-ice system. The snow would sublime away in a few hours by heating the dish panels a couple of degrees C above the ambient temperature.

With the de-ice on, there was no noticeable effect on the pointing accuracy for the Rx86 and Rx230 observations. Although we do not normally expect to operate the telescope in snow conditions, the operation mode with de-ice on can be helpful in time-critical observations, e.g., for VLBI campaigns where the observations cannot be postponed.

### 8.5 Antenna Tilting

There are 3 tiltmeter pairs placed on the GLT antenna:

- Tiltmeter Pair 1: located in the yoke arm (above the AZ bearing), with +X is toward the front of the antenna and Y-axis parallel to the elevation axis with +Y toward the right side when looking into the dish.
- Tiltmeter Pair 2: located in the support cone, with X towards the West and Y towards the South.
- Tiltmeter Pair 3: located in the steel foundation, with X towards the East and Y towards the North.

The long-term tiltmeter recordings show a correspondence between the tiltmeter readings and the ambient temperature. The large-scale variations in the telescope tilt on timescales of months and years due to the changing seasons at Pituffik, resulting in long-term drifts in the GLT pointing. These long-term pointing drifts are typically less than 10" peak-to-peak for most of the tiltmeter readings, except Tiltmeter 3 in the X direction, which has 30" peak-to-peak variations over the measured timespan. The source of these variations in the telescope tilt is still being investigated. To compensate for this, we conduct pointing measurements immediately before scientific observing campaigns to update the pointing model, particularly for the tilt-related terms.

Most importantly, the intra and inter-day variations of the tilt are much more minor (less than 5 arc-second peak-to-peak, corresponding to less than 2 arc-second RMS). This indicates that this temperature dependence of the pointing on short timescales does not affect science observations on a day-to-day and weekly basis.

To investigate the effect of wind gusts on the stability of the antenna mount, we recorded tiltmeter data in fast mode while simultaneously recording the wind speed at the rate of 2 Hz. The data were taken on 31 March 2021 when we had storm conditions with wind speeds exceeding 20 m/s. The antenna was stowed in the survival stow position of AZ=5.8 and EL=15 deg. Significant changes in Tiltmeter Pair 1 are correlated with the high wind speed. The upper part of the antenna (above the azimuth bearing) has a change in tilt by about 17 arcseconds. The base part also shows changes of about half that amount. Fortunately, under nominal high-speed conditions (up to 5 m/s gusts), the variation in all tiltmeters appears below 2 arcseconds. The changes with very high wind gusts may be purely elastic, but more data are needed to confirm this.

### 8.6 Aperture Efficiency and Surface Accuracy

With a 12 dB edge taper illumination, the GLT's aperture efficiency is capped at 78%. With the pointing capability established, we measured the aperture efficiency, $\eta_A$. The technique follows those described in Mangum 1993 [41]. Firstly, we derived the main-beam efficiency by observing a planet with known temperature and size and the forward spillover efficiency by observing the Moon. With these numbers, we found the effective aperture of the antenna, which was then used to derive the $\eta_A$.

The first measurement was taken in March 2018, and the results show ~ 22 % aperture efficiency at 230 GHz, and ~ 48 % at 86 GHz. These results reflected the poor surface accuracy of the antenna dish. The measured aperture efficiency implied a surface error of 150-200 um RMS, based on Ruze's gain-degradation factor at 86 GHz[16], assuming that the ohmic loss is negligible. After the first panel adjustment in 2018 Summer, the aperture efficiency measured in January 2019 yielded a result of 60% for the 230 GHz. It fitted to a surface error of 50 μm RMS in Ruze's equation.

After the panel adjustments in 2019 Summer, we conducted the 230 GHz aperture efficiency measurements in Mar. 2020 and 86 GHz in Nov. 2020. The 230 GHz result was 66%, and the 86 GHz was 75%. The additional photogrammetry and the surface adjustments improved the surface accuracy, and the current efficiencies matched well with the surface accuracy of 40 μm. A detailed account of the aperture efficiency measurement of the GLT can be found in Ref. 39.

Table 7 compiles a timeline of the surface accuracy of the primary dish derived from various measurements, supplemented with the measurements' time and nominal ambient temperature.

### 9. Calibration and Sensitivity

As part of the global array, we follow the calibration steps outlined by the EHT collaboration [2], which relates the correlated flux density in terms of each telescope's system equivalent flux density (SEFD), which can be expressed as

$$SEFD = \frac{T^*_{sys}}{DPFU \times g_E} \quad , \tag{5}$$

where $T^*_{sys}$ is the effective system noise temperature corrected for atmospheric attenuation, $g_E$ is the elevation-dependent gain curve of the telescope, and DPFU the conversion factor of the noise temperature per flux density unit (DPFU).

---

[16] For example, John D. Kraus and Ronald J. Marhefka, Antennas, 3rd edition, McGrawHill, p. 667.

Table 7
The historical data of the surface accuracy of the GLT primary dish.

| Time & Ambient Temperature | Surface Accuracy derived from various measurements (RMS) | |
|---|---|---|
| 2017 Apr @ -15 °C | 40 μm | Laser tracking, pre-assembly |
| 2018 Mar @ -25 °C | 150-200 μm | $\eta_{A,86} = 22\%$, $\eta_{A,230} = 48\%$ |
| 2018 Aug @ 3 °C | 170 μm / 17 μm | Photogrammetry before and after panel adjustment |
| 2019 Jan @ -25 °C | 50 μm | $\eta_{A,230} = 60\%$ |
| 2019 Apr @ -15 °C | 26 μm | After panel adjustment based on photogrammetry |
| 2020 Mar @ -25 °C | 40 μm | $\eta_{A,230} = 66\%$ |
| 2020 Nov @ -25°C | 40 μm | $\eta_{A,86} = 75\%$ |

DPFU can be expressed as:

$$DPFU = \frac{A_e}{2k_B} \quad , \tag{6}$$

where the effective aperture, $A_e$, relates to the aperture efficiency times the physical size of the dish, and $k_B$ is the Boltzmann constant.

The antenna system noise temperature can be expressed, in good approximation, in the following form:

$$T_{sys} = T_{rx} + T_{atm} \cdot \eta_l \cdot (1 - e^{-\tau}) + T_{amb} \cdot (1 - \eta_l) \quad . \tag{7}$$

The first term on the right-hand side of Equation (7), $T_{rx}$, is the receiver noise temperature. The second term represents the noise coming from an atmosphere with a temperature $T_{atm}$ and the opacity τ, moderated by the antenna's forward efficiency, $\eta_l$. The third term considers the noise contribution from the ambiance, excluding the forward direction, at a temperature of $T_{amb}$.

We differentiate $T_{atm}$ and $T_{amb}$ because we keep the receiver cabin at a set temperature of 15-20 °C for the electronic systems to function properly. In contrast, $T_{atm}$ is typically much colder during the Arctic winter. Thus, the effective system temperature, should it be measured with a perfect antenna at a location above the atmosphere, is:

$$T^*_{sys} = \frac{e^\tau}{\eta_l} T_{sys} \quad . \tag{8}$$

The $T^*_{sys}$ is derived in the following steps. First, we measure the receiver power output with the antenna looking at the cold sky. The output power, $P_{sky}$, relates to the $T_{sys}$ via a gain factor of the antenna system. We also measure the power from the ambiance by placing the calibration absorber in front of the receiver (see Section 3.5). The ambient load, $P_{amb}$, is proportional to $T_{amb} + T_{rx}$, also via the gain factor as the sky observation. With these two measurements, the effective system temperature can be expressed as:

$$T^*_{sys} = \gamma \cdot T_{atm} \left(1 + e^\tau \frac{T_{amb} - T_{atm}}{T_{atm}}\right) \quad , \tag{9}$$

where

$$\gamma = \frac{P_{sky}}{P_{amb} - P_{sky}} \quad .$$

A sensor installed at the calibration absorber monitors the $T_{amb}$. The effective atmospheric temperature, $T_{atm}$, is monitored continuously by an on-site weather station.

Due to the scarcity of cryogen at PSB, we resolve using sky-dipping to measure the receiver noise temperature, $T_{rx}$. However, we brought in liquid nitrogen to check the receiver performance at least once a year.

The atmospheric opacity is measured with an on-site tipping radiometer, operating at 225 GHz and taking zenith opacity data every 10 min. Occasionally when the radiometer was out of commission, we measured the opacity with the GLT as the sky-dipper. Since 2018, we have accumulated five years of results [42]. The PAB site shows an evident seasonal variation with average opacity lower by a factor of two during winter compared with summer, similar to Greenland Summit [8]. The 25%, 50%, and 75% quartiles of the 225 GHz opacity

Table 8
Comparisons of the GLT performance between 2018 and 2021

|                        | RX86    |         | Rx230   |         |
| ---------------------- | ------- | ------- | ------- | ------- |
| Year                   | 2018    | 2021    | 2018    | 2021    |
| Aperture efficiency    | ~48%    | 75% ± 1% | ~22%   | 66% ± 1% |
| DPFU (K/Jy)            | 0.01966 | 0.03072 | 0.00901 | 0.02703 |
| SEFD (Jy)              | 12,869  | 8,236   | 62,925  | 20,975  |

Note: The DPFU uncertainties include a nominal 5% error in the estimate of the aperture efficiencies. The $T_{sys}$ used to derive the SEFD is 253K for 86 GHz, and 567K for 230 GHz, obtained during the 2021 GMVA and EHT campaigns.

during the winter months of November through April are 0.14, 0.17, and 0.22, respectively. These values are three times larger than those at the Greenland Summit, and the astronomical observing efficiency in this band is about an order of magnitude better at the Summit than at the PAB.

The value of the DPFU is derived from the aperture efficiency data. Given that the aperture efficiencies are obtained based on observations of sources (such as planets) with available sizes and temperatures, the DPFUs should also include 5% uncertainties of the flux density model for the planets observed.

We need more data to model the variation of the antenna efficiencies over elevation. This is because the telescope is located at such a far north latitude. There are no calibrator planets or other targets the telescope can track over an extensive range of elevations to measure. However, the typical VLBI science targets for the GLT always lie in low elevations and do not significantly change elevation during observations. Therefore, the elevation-dependent gains can be assumed constant in these VLBI science observations, compared to the other uncertainties in the antenna efficiency measurements. The same assumptions have been adopted for the South Pole Telescope (SPT) in EHT data [43].

We are exploring the variation of the antenna efficiencies as a function of time of day, ambient temperature, and elevation (where possible) to better characterize the telescope gains for future EHT campaigns. Although calibrator planets appear above the horizon for a limited time, we aim to collect such data over many observing seasons, so that sufficient statistics can be obtained for these types of analyses. Table 8 summarizes the GLT's calibration data obtained so far.

Since the EHT dress rehearsal in Jan 2018 and scientific observations in April 2018, the GLT has participated in many VLBI observations as part of the EHT, GMVA, and EAVN-Hi (East-Asia VLBI Network, High-frequency) network. Among all the activities, fringes were detected in the 2018 EHT 14 VLBI observations, no fringes in two, and two more science runs yet to be correlated. Figure 11 shows the correlated fringe between the GLT and ALMA from the test campaign in Jan 2018. It is unclear why there was no fringe, although it occurred during the first two years of our operations. Among those with fringe, detection is in all the receiver bands, including the 345 GHz. From January 2019, the VLBI backends were upgraded from a recording bandwidth of 32 Gbps to 64 Gbps, to match that of the upgraded EHT.

## 10. Summary

PSB is the staging site for deploying the GLT to Summit Station. Although the site is at sea level, its cold winter sky still provides us with realistic conditions for testing the instrumentation in the field and, most of all, the opportunities for conducting successful scientific observations. Through these exercises, we have learned lessons about telescope operations in the Arctic region. Due to the travel restriction of the COVID-19 pandemic, we have developed the capability and the protocols to operate the GLT remotely. We successfully carried out all-sky-pointing observations remotely in October and November 2020. We are continually improving the remote operational capabilities of the telescope, which are essential for the operation at Summit Station.

Before the summit deployment, we will further refine the telescope's functions for scientific operations; such as the nutating capability and the holography program. We will devote a significant amount of effort to modeling the temperature variation effect on the dish surface accuracy. In the meantime, we plan to expand the suite of instruments to include a sub-mm camera and a multi-pixel heterodyne receiver. Work is underway to create physical space in the receiver cabin for instruments designed for Nasmyth's focus. We plan to implement a beam-selecting mechanism to redirect the incoming signal to one of the receivers. For the VLBI receiver, we will implement new optics to enable simultaneous multi-beam operations. We will need additional LO signal sources to support such operations.

Beyond the Pituffik operations, the transportation of the GLT over the ice sheet and the construction of the observatory infrastructure require extraordinary engineering and unconventional expertise. We are working closely with the experts of the Office of the Polar Program (NSF), Battelle Arctic Research Operations, and the Cold Regions Research and Engineering Laboratory (CRREL) to develop a comprehensive deployment plan. With the participation of the Danish collaboration, we can tap into the resource and experience of the East Greenland Ice-core Project (EastGRIP), which has extensive working history on the Greenland ice sheet. The deployment and the observatory construction are a project of their own, and the subject matter is beyond the scope of this paper. They will be described elsewhere.

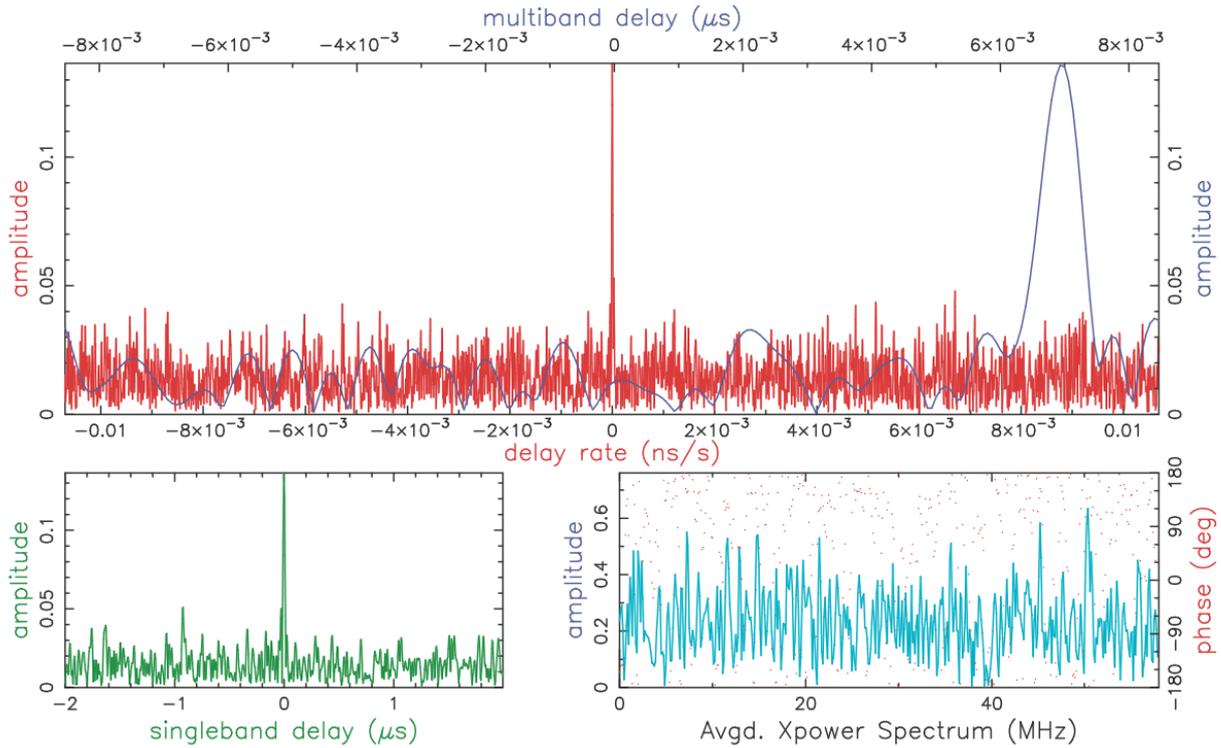

Figure 11. Results of the fringe fitting towards 3C 279 between GLT and phase-up the ALMA. The data were taken as a part of the test VLBI observations in the framework of EHT on Jan 28$^{th}$, 2018, and correlated at the MIT Haystack Observatory in Westford, Massachusetts, USA. The top panel shows the fringe plot in multi-band delay (blue) and delay rate (red); the bottom-left panel is the results of the fringe fitting in single-band delay, while the bottom-right panel presents the amplitude (blue) and phase (red) of the bandpass character of the visibility between phased ALMA and GLT.

This paper summarizes the construction, commissioning, and current status of the GLT since the telescope arrived in Greenland. It also describes the telescope's system characteristics and capability as a VLBI station operating in sub-mm wavelengths. The GLT has participated in many VLBI observations since 2018, and the scientific outcome derived from the data will soon be available. Yet our goal is to establish the GLT on a high site in Greenland. Summit Station is a year-round operational camp at a premier location for observing in short sub-mm wavelengths, and it also provides the critical baselines for global VLBI observations. The GLT is ready and capable of exploiting these distinctive potentials of Greenlandic ice sheet.

**Acknowledgment:**

The authors would like to thank Claus Gonnsen, Ane-Louise Ivik, Chen-Yu Yu, Lupin C. Lin, Zheng Meyer-Zhao, Alex Alladi, Aaron Faber, and Nicolas Pradel for their expertise and assistance to the development of the GLT project. The funding for the GLT is partially supported in ASIAA by the Academia Sinica and by the Ministry of Science and Technology, MOST funding codes: 99-2119-M-001-002-MY4, 103-2119-M-001-010-MY2, and 106-2119-M-001-013, for Taiwan's participation in the ALMA-NA project. The realization of the GLT provides a unique baseline for the ALMA's VLBI capability. The GLT project is also partially supported at the SAO by the Smithsonian Institution. Support for the Hydrogen Maser frequency standard used for VLBI at the Greenland Telescope was provided through an award to SAO from the Gordon and Betty Moore Foundation (GBMF-5278). The GLT project is thanks to the strong support from the NCIST and the James Clerk Maxwell Telescope under the management of the East Asia Observatory. We also thank IRAM and the National Astronomical Observatory of Japan for their support in the receiver instrumentation, the US National Science Foundation, Office of Polar Programs for support in logistics, and the United States Air Force, 821$^{st}$ Air Base Group, Thule Air Base, Greenland for the use of the site and access to the Base and logistics chain. The project appreciates the development work done by Vertex Antennentechnik GmbH and ADS international. We would also like to thank Atunas, an outdoor wear company in Taiwan, for the arctic clothing. Lastly, we thank the National Radio Astronomy Observatory and MIT Haystack Observatory for their support in the acquisition of the ALMA-NA prototype to repurpose it for Greenland deployment. The GLT project acknowledges the privilege of operating on the lands of Greenland and strives to work with the local government and the people of Greenland to exploit the scientific and educational potentials of this project. The GLT project pledges to respect and preserve the environment we work in. The project will be responsible for removing our instrumentation and restoring the environment to its natural state at the end of this project.